\documentclass[a4paper,fleqn]{cas-sc}

\usepackage[authoryear,longnamesfirst]{natbib}
\usepackage[linesnumbered,algoruled,boxed,lined]{algorithm2e}
\usepackage{graphicx}
\usepackage{multicol}
\usepackage{multirow}
\usepackage{tabularx}

\def\tsc#1{\csdef{#1}{\textsc{\lowercase{#1}}\xspace}}
\tsc{WGM}
\tsc{QE}

\begin{document}
%
%
%
\title [mode = title]{Secure and Energy-efficient Unmanned Aerial Vehicle-enabled Visible Light Communication via A Multi-objective Optimization Approach}  
%
%
%

\author[1]{Lingling Liu}
\author[1]{Aimin Wang}
\author[1]{Jing~Wu}
\ead{wujing_jlu@hotmail.com}

\cormark[1]

\address[1]{College of Computer Science and Technology, Jilin University, Changchun 130012, China}
\address[2]{Key Laboratory of Symbolic Computation and Knowledge Engineering of Ministry of Education, Jilin University, Changchun 130012, China}

\author[1]{Jiao Lu}
\author[1]{Jiahui Li}
\author[1,2]{Geng Sun}

\cortext[1]{Corresponding author}


\begin{abstract}
In this research, a unique approach to provide communication service for terrestrial receivers via using unmanned aerial vehicle-enabled visible light communication is investigated. Specifically, we take into account a unmanned aerial vehicle-enabled visible light communication scenario with multiplex transmitters, multiplex receivers, and a single eavesdropper, each of which is equipped with a single photodetector. Then, a unmanned aerial vehicle deployment multi-objective optimization problem is formulated to simultaneously make the optical power received by receiving surface more uniform, minimize the amount of information collected by a eavesdropper, and minimize the energy consumption of unmanned aerial vehicles, while the locations and transmission power of unmanned aerial vehicles are simultaneously optimized under certain constraints. Since the formulated unmanned aerial vehicle deployment multi-objective optimization problem is complex and nonlinear, it is challenging to be tackled by using conventional methods. For the purpose of solving the problem, a multi-objective evolutionary algorithm based on decomposition with chaos initiation and crossover mutation is proposed. Simulation outcomes show that the proposed approach is superior to other approaches, and is efficient at improving the security and energy efficiency of visible light communication system.
\end{abstract}




\maketitle

%
%
\section{Introduction}
\label{sec1}
\label{introduction}
\par Visible light communication (VLC) is particularly suitable for such scenarios that require both illumination and communication services~\citep{lu2022Optical}, relying on light-emitting diodes (LEDs) and photodiodes (PDs) for signal transmission, and is a potential technique for short-range wireless communication link using lighting infrastructure~\citep{mostafa2015physical}. On the one hand, due to line-of-sight (LoS) dominates the VLC link, and the opaque surface will hinder the transmission of light waves, this link will be interfered with limited or no internetwork interference. On the other hand, although VLC channels are LoS propagation and have better signal transmission, there is no optical fiber or any kind of waveguide to assist transmission. Comparing to radio frequency (RF) channels, VLC still has broadcast characteristics, which makes VLC links inherently vulnerable to unintentional access or unauthorized eavesdropping by users in some physical areas illuminated by the data transmitters~\citep{mostafa2015physical}. As a result, it is crucial to be able to safely share sensitive information when there are adversaries present, since they might try to launch various assaults to get unauthorized access and modify some information, or even disrupt regular information flows~\citep{shiu2011physical}, especially in some public areas, such as classrooms, shopping malls and libraries.

\par Therefore, security and confidentiality in VLC are becoming a close concern of users and network administrators. To this end, the upper-layer encryption technology is the commonly used security technology in VLC~\citep{liang2020physical}, in which this network can be regarded as secure as long as the storage capacity and the computing power of the potential eavesdroppers are kept within a certain range. However, although this technique can also be used in VLC system, it is not recommended because it requires higher computing performance of the receivers to accurately decipher the obtained information, and this technique depends on a trustable key transmission mechanism. 

\par Physical layer security (PLS), which can act as the initial line of defense against threats, has long been a potential research focus, while complementing traditional encryption techniques,~\citep{mukherjee2014principles},~\citep{abbas2018opportunistic},~\citep{hu2017cooperative},~\citep{lee2018adaptive}. PLS refers to confusing the potential unauthorized recipients by employing the characteristics of the channel and impairing their ability to infer information through specially-designed signaling and/or coding schemes, so that enhancing the system security~\citep{mukherjee2014principles},~\citep{liang2009information}. These results are based on the confidential information theory features, that can be traced back to some of the outstanding work of Claude Shannon on mathematical theory of communication~\citep{shannon1949communication}. Unlike the researches of Shannon mainly focusing on the symmetric key encryption systems, those of Aaron Wyner in~\citep{wyner1975wire} mainly focus on the characteristics of the wiretapper channel model, that is, the communication channel itself can transmit secrets without receiving the key shared by the transmitters and receivers. Furthermore, it is demonstrated in~\citep{wyner1975wire} that, theoretically, it is feasible to achieve totally secure communications if the wireless channel of authorized users is stronger than that of eavesdropper. But if the wireless channel authorized users is subpar, a extremely low or even zero secure transmission rate is possible~\citep{lan2020achievable}.

\par The authors in~\citep{kumar2020pls} investigate the PLS in an indoor environment to prevent eavesdropper attacking by imposing limits on secrecy rate. Specifically, the VLC and RF technologies are employed together, in which the former is used as the primary technology and the latter will be used when the primary technology cannot meet the imposed restrictions. Then, in order to increase secrecy rate, the secure transmission protocol based on artificial noise (AN) is proposed, in which AN is deliberately introduced to interfere eavesdroppers in VLC network~\citep{peng2021physical},~\citep{wang2017artificial},~\citep{goel2008guaranteeing}. The PLS for a multi-input single-output (MISO) non-orthogonal multiple access (NOMA) VLC system is studied in~\citep{peng2021physical}, in which a precoding method is proposed to increase the secrecy capacity of system by reducing the rate of collusive eavesdroppers. Specifically, to further minimize the average rate of PLS of MISO-NOMA-VLC system when it is intercepted by collusive eavesdroppers, the AN is introduced in precoding scheme. Generally speaking, multi-antenna transmission technologies are frequently utilized with AN as a different approach to increase the secrecy rate by utilizing spatial degree of freedom. For example, in~\citep{wang2017artificial}, the authors study PLS in multi-antenna small-cell networks, in which each base station (BS) applies AN to assist transmission to increase the secrecy performance of network.

\par In order to ensure the channel security, the transmitter will use multiple-input multiple-output (MIMO) to generate information signals and its own friendly jamming (FJ), and the legitimate receiver will use MIMO zero-forcing to remove this jamming. However, a ``vulnerable area" will be left around each legitimate receiver, which is easily attacked by illegal eavesdroppers, therefore the authors in~\citep{akgun2016exploiting} solve this problem by enhancing the Tx-based FJ (TxFJ) with Rx-based FJ (RxFJ) generated by each legitimate receiver. The eavesdroppers in the above examples are single antenna, and there are still some examples of multi-antenna eavesdroppers that have been widely studied. In the existence of passive eavesdroppers with multi-antenna, Goel et al.~\citep{goel2008guaranteeing} investigate the secret communication between two nodes in fading wireless channel, in which the secrecy of communication is achieved. In~\citep{chu2015secrecy}, in the presence of multi-antenna eavesdroppers, the problems of maximizing secrecy rate and minimizing power of MISO secure channel are studied.

\par Then the beamforming vectors are also often used to deal with the problems in wireless communication networks~\citep{sun2021time} and achieve the PLS~\citep{al2018physical},~\citep{liu2020beamforming}. For instance, the authors in~\citep{al2018physical} design RF-and-VLC-based beamforming vectors for maximizing the achievable secrecy capacity (SC), and the power minimization problem is solved by using these vectors under the constraints of the required SC.

\par Nevertheless, VLC system usually installs LED lights on the wall or ceiling, and it cannot move once installed, which is inconvenient in reality. With the development of automation and semi-automation vehicles, such as unmanned aerial vehicles (UAVs) and robots, have attracted great attention in various applications. The increasingly widespread use of LEDs for communication in UAVs provides possibilities for VLC applications, like monitoring, data collection of Internet of thing devices, uploading traffic data from BSs, and increasing service quality for users~\citep{eltokhey2021uav}. Therefore, in this work, we adopt UAVs to assist VLC to provide service for receiving surface. Moreover, in VLC networks, the applications based on UAV have received particular interests. In~\citep{pham2020sum}, in an effort to increase the total rate of all users in UAV-enabled VLC, a joint power allocation and UAV deployment problem is developed. The authors in~\citep{peer20223} propose the energy and user mobility aware 3D deployment of VLC-enabled UAV-base station (UBv) to maximize user coverage while ensuring fairness. By optimizing the locations of UAVs and association, the power consumption of UAVs is reduced~\citep{yang2019power}, which is further extended in~\citep{wang2020deep} and solved by using deep learning.

\par However, there are still some problems in UAV-enabled VLC system. The quality of communication on the same plane varies as a result of the broadcast features of VLC. Therefore, so as to ensure that the received optical power on the same receiving surface is uniform, which can guarantee that the receiver at any place can have access to the network more fairly, it is required to improve communication quality. Moreover, if only one UAV is used to provide communication in this system, there will be uneven services on the receiving surface. Therefore, in this paper, we will use multiple UAVs and find a group of best hovering locations to achieve this target. In addition, in the system with an eavesdropper, the security or secrecy of this system also needs to be considered. Finally, the UAVs will consume corresponding energy in the process of finding optimal hovering locations, which will limit their ability to communicate and decrease their ability to provide continuous services over an extended period of time~\citep{mozaffari2019tutorial}. Accordingly, we will comprehensively consider these problems since there is a trade-off relationship between them.

\par As a result, we focus on investigating the UAV-enabled VLC system, when an eavesdropper is presented in this work, in which multiple UAVs are adopted to provide services for receiving surface. In contrast to past efforts, which only considered UAV communication, security, and energy consumption separately, we provide a multi-objective optimization strategy to simultaneously optimize several objectives for thoroughly resolving these challenges. The primary contributions of this study are summed up as follows:

\begin{itemize}
	\item We take into account such a UAV-enabled VLC case, in which numerous UAVs are deployed to provide communication for receiving surface. Then, a UAV deployment multi-objective optimization problem (UAVDMOP) of UAV-enabled VLC system is formulated to collectively make the received optical power uniform on the receiving surface, minimize the amount of information the eavesdropper gathers, and minimize the overall energy consumption of UAVs, while adhering to certain restrictions.
	
	\item The formulated UAVDMOP is sophisticated and nonlinear, making it challenging to be solved by using classical methods. In order to tackle this issue, we propose a multi-objective evolutionary algorithm based on decomposition with chaos initialization and crossover mutation (MOEA/D-CICM). A chaos initialization technique is used by MOEA/D-CICM to improve the performance of initial solutions. Moreover, a crossover mutation operator is introduced to make the algorithm have better exploration and exploitation abilities in dealing with the UAVDMOP.
	
	\item Simulations are carried out to evaluate the performance and effectiveness of the proposed method against other methods. Results show that the proposed MOEA/D-CICM has the ability to make the optical power received by the receiving surface more uniform, minimize the amount of information received by the eavesdropper, in addition to minimize the energy consumption of UAVs.
\end{itemize}

\par The rest of this work is structured as follows. The system model and preliminaries are presented in Section \ref{Models and preliminaries}. In Section \ref{Problem formulation}, the UAVDMOP of UAV-enabled VLC is formulated. Section \ref{The proposed MOEA/D-CICM} proposes the algorithm. The simulation results are shown in Section \ref{Simulation results}. Finally, Section \ref{Conclusion} summarizes the conclusion.

%
%
\section{Models and Preliminaries}
\label{Models and preliminaries}

%
%
\begin{figure}[!htbp]
	\centering
	\includegraphics[width=3in]{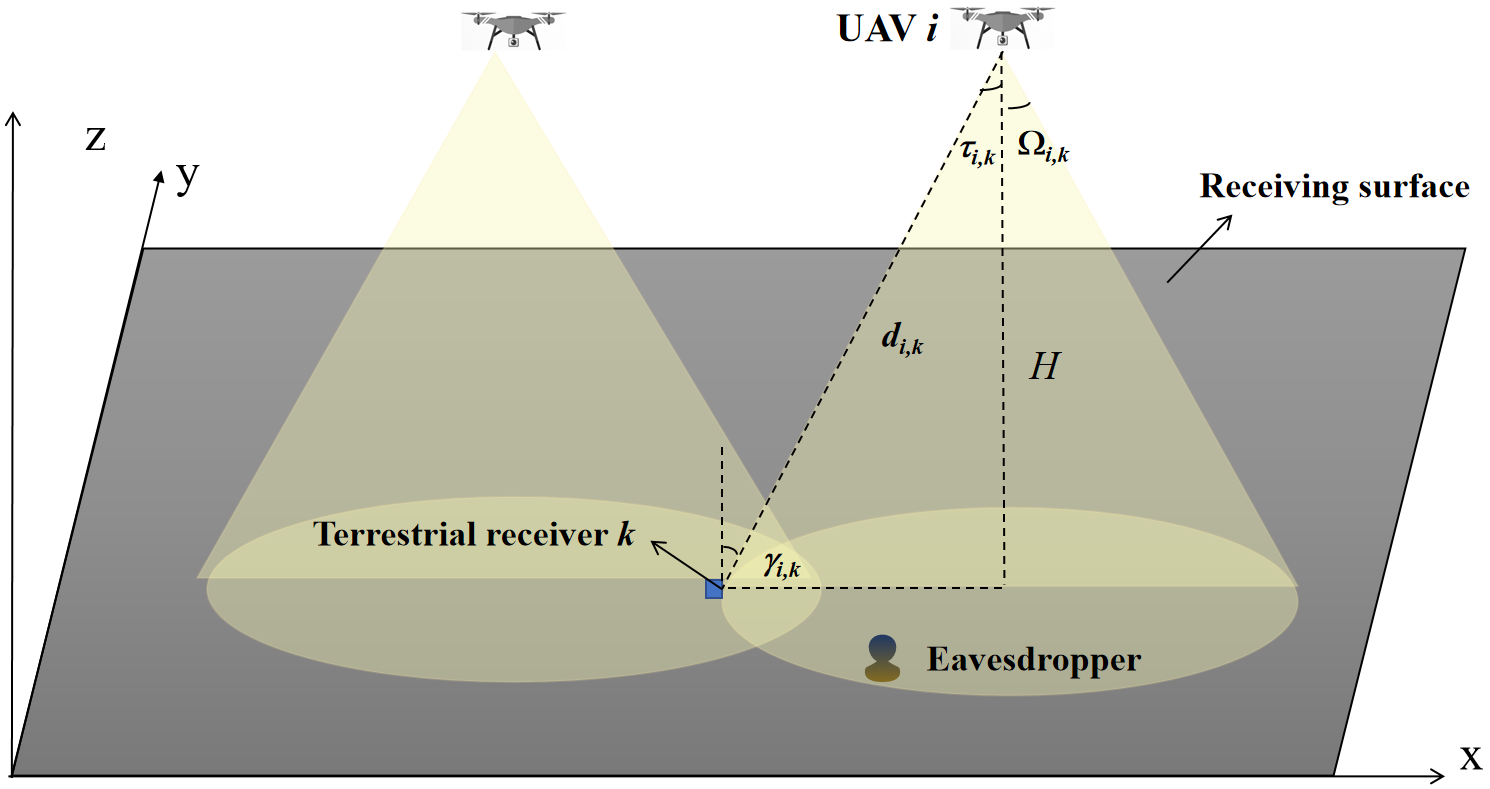}
	\caption{The UAV-enabled VLC network consisting of UAVs and receiving surface.}
	\label{UAV user}
\end{figure}
\subsection{VLC channel model}
\par We take into account a UAV-enabled VLC network shown in Fig. \ref{UAV user}. In this network, a group of rotary-wing UAVs $\mathcal{U}$ equipped with LEDs are used to provide communication for target receiving surface. Moreover, a set of receivers $\mathcal{N}$ on the target area can be selected as the samples and used to evaluate the communication quality of the receiving surface. Therefore, the sets of UAVs and receivers can be respectively denoted as $\mathcal{U} = \{1, ..., U\}$ and $\mathcal{N}=\{1, ..., N\}$. It is assumed that the LEDs in VLC system follow Lambert radiation pattern. Due to LoS link is the primary element of an optical channel~\citep{2019Camera}, we only consider LoS channel. Therefore, the channel gain $h_{i,k}$ of VLC link between UAV $i$ and receiver $k$ is designed by~\citep{komine2004fundamental}:

\begin{equation}	
	\label{eq:channel gain}
	h_{i,k}=\left\{
	\begin{array}{l} 
		\frac{(m+1) A_r}{2 \pi d_{i,k}}g(\psi_{i,k})\cos^{m}(\phi_{i,k})\cos(\psi_{i,k}), 0 \leq \psi_{i,k} \leq \Psi_{c}, \\ 
		0, \quad \quad \quad \quad \quad \quad \quad \quad \quad \quad \quad \quad \quad \ \ \psi_{i,k} > \Psi_{c}.
	\end{array}\right.				
\end{equation}

\noindent where the detector area is denoted by $A_r$. The receiver field of vision (FOV) semi-angle is indicated by $\Psi_{c}$, and $m=-\ln2/\ln(\cos\Phi_{1/2})$ represents the Lambert index, in which $\Phi_{1/2}$ is the transmitter semi-angle. In addition, the distance between UAV $i$ and receiver $k$ can be denoted by $d_{i,k}$, and the angles of incidence and irradiance are expressed by $\psi$ and $\phi$, respectively. The gain of the optical concentrator $g(\psi_{i,k})$ is expressed as~\citep{komine2004fundamental}:

\begin{equation}		
    g(\psi_{i,k})=\left\{
    \begin{array}{l} 
    \frac{n_{r}^2}{\sin^2(\Psi_{c})}, \ \psi_{i,k} < \Psi_{c}, \\ 
    0, \quad \quad \quad \psi_{i,k} \geq \Psi_{c}.
    \end{array}\right.	
\end{equation}

\noindent where $n_{r}$ represents the refractive index. It is noted that these angles are obtained under the assumption that, the directions of transmitter and receiver are vertically downward and upward, respectively. It is assumed that all receivers are on the ground (i.e., the altitude of receivers are zero) and the UAVs are moving at a fixed altitude of $H$. The positions of UAVs and receivers are denoted by $(x_{i}^U, y_{i}^U, H)$, $i = 1, 2, ..., U$, and $(x_{k}^G, y_{k}^G, 0)$, $k = 1, 2, ..., N$. In addition to the distance between UAV $i$ and ground receiver $k$ is calculated as follows:

\begin{equation}
	d_{i,k}=\sqrt{(x_{i}^U - x_{k}^G)^2 + (y_{i}^U - y_{k}^G)^2 + H^2},
\end{equation}

\par Moreover, the angles values of irradiance and incidence can be expressed as $\cos\psi_{i,k}=\cos\phi_{i,k} = H/d_{i,k}$.

%
%
\subsection{Achievable information rate}
	
\par Since each receiver on the ground can receive the optical signal transmitted by the corresponding UAVs within a certain range, for receiver $k$ located at ($x_k^G, y_k^G$), the achievable information rate $R_{i,k}$ from UAV $i$ can be given by:
	
\begin{equation} 
    \label{achievable information rate}
    R_{i,k} = \frac{1}{2}\log_2 \left(1+\frac{e}{2 \pi}
    \frac{ ( P_i g_{i,k})^2}{\sum_{r = 1, r \ne i}^{N} (P_r g_{r,k} )^2 +\sigma_w} \right),
\end{equation}	
	
\noindent where $P_i$ and $\sigma_w$ represent the transmission power of UAV $i$ and the standard deviation of the additive white Gaussian noise (AWGN). Gaussian noise can be used to simulate the impact of ambient light since the UAV-enabled LEDs are usually deployed in some places with poor light \cite{komine2004fundamental}.
	
%
%
\subsection{Power adjustment factor of LED}
	
\par In terms of the receiver $k$ covered by UAVs, the total received optical power can be calculated as:
	
\begin{equation} 
\label{eq:Received Signal}
    P_{r}(k)=\sum_{i=1} ^{C_{o}} {g_{i,k}} {P_{i}}, k \in K,
\end{equation}
	
\noindent where $C_o$ is the number of UAVs that serves receiver $k$.
	
\par It can be seen from Eq. \eqref{eq:Received Signal} that the received optical power depends on the transmission power of UAV $P_i$ and the channel gain $g_{i,k}$ between UAV $i$ and receiver $k$. Thus, we introduce a power adjustment factor for each LED to improve the power distribution so that assuring each receiver $k$ can receive almost the same power, and it can be computed as $P_{k}=\omega_{i,k}P_{i}$. Thus, for a receiver $k$, the received optical power can be rewritten as:
	
\begin{equation}
    P_{r}(L_{k})=\sum_{i=1}^{C_{o}} \omega_{i,k} g_{i,k} P_{i},
\end{equation}

\subsection{Energy consumption model of UAV}
\par The energy consumption of a UAV is mostly composed of the energy consumption related to communication and propulsion \cite{2019Energy}. In contrast to the latter, which tends to assist the upward and forward movement of UAVs, the former is the energy consumption related to the circuit, signal reception, and processing. Additionally, the power to communicate while hovering might be regarded as a constant. The following equation can be used to estimate the propulsion power consumption of a rotary-wing UAV traveling at a speed $V$:
\vspace{-0.3cm}
\begin{equation}
	\begin{aligned}
		\label{horizontal model}
		\\P\left ( V\right )=&\underset{\textbf{blade\ profile}}{\underbrace{P_{0}\left ( 1+\frac{3V^{2}}{U_{tip}^{2}}\right )}}+\\&\underset{\textbf{induced}}{\underbrace{P_{i}\left ( \sqrt{1+\frac{V^{4}}{4v_{0}^{4}}}-\frac{V^{2}}{2v_{0}^{2}}\right )^{1/2}}}
		+\underset{\textbf{parasite}}{\underbrace{\frac{1}{2}d_{0}\rho sAV^{3}}},
	\end{aligned}
\end{equation}

\noindent where $P_0$ and $P_i$ represent the power of the blade profile and the induced when hovering, respectively. The tip speed of the rotor blade and mean rotor induced velocity when hovering are denoted by $U_{tip}$ and $v_0$, respectively. The terms $d_0$ and $s$ stand for the fuselage drag ratio and the rotor solidity, respectively. The air density and rotor disc area are denoted by $\rho$ and $A$, respectively.

\par It is worth mentioning that in this work, since it only makes up a small portion of the total operation time during the maneuvering duration of the UAVs, we do not consider the energy consumption incurred by the acceleration and deceleration while the UAVs fly horizontally.

%
%
\section{Problem formulation} 
\label{Problem formulation}

\par In the considered scenario, the selected receivers are randomly distributed on ground and an eavesdropper is also existed somewhere on the system, and UAVs have some distance with them, which will result in different service quality on various receivers. This means that in some originally dim places, the lighting is still poor after using a UAV. Moreover, this will also make most of the information transmitted by UAV likely to be received by the eavesdropper, so that affecting the security of communication system. However, the energy of a UAV is usually limited, so it is unrealistic to continuously transmit information and provide more light to receivers by improving the transmission power of a UAV, and this does not prevent the eavesdropper from eavesdropping on information. Therefore, it may be a better strategy by using multiple UAVs and optimizing their locations for providing communication for terrestrial multiple receivers, so that improving the optical power level in safer places, meanwhile minimizing the amount of information eavesdropped by the eavesdropper. 

\par Nevertheless, there may be a number of challenges when deploying UAVs in such a system. First, although the UAVs deployed at higher altitude can provide communication for a wider range of surface, it will make the received optical power uneven in most of receivers, and even some blocked parts are still dim. Second, in the system with eavesdropper, UAVs have the advantage of providing communication for more places, while it is still a difficult problem to be solved in ensuring security of the system. Finally, UAVs will increase the mobile energy consumption in the process of deployment, which makes it difficult for UAVs to continuously provide communication for receiving surface.

\par Accordingly, as there are trade-offs between the aforementioned issues, they must be resolved thoroughly. Therefore, by estimating the ultimate hovering locations and transmission power of UAVs, we formulate a UAVDMOP in UAV-enabled VLC system, and the associated optimization objectives are detailed below.

\textbf{\emph{1) Optimization objective 1}}: In this work, we select a set of sampling points from receiving surface as receivers, so as to ensure that the optical power received by the receiving plane is more uniform. The degree of sample volatility can be accurately reflected by the variance. In case of the same data samples, a larger variance indicates a larger fluctuation of data, which means the more unstable data distribution. For $N$ samples on receiving plane, following are the formulas for calculating the average received optical power:

\begin{equation}
	Ave(P_{r})=\sum_{j=1}^N P_{r}(L_{k})/N,
\end{equation}

\par Therefore, the objective function to make the received optical power on receiving surface more uniform can be designed as follows:

\begin{equation}
	f_1(\mathbb{X}, \mathbb{Y}, \mathbb{Z}, \mathbb{P}) = \frac{1}{N}\sum_{j=0}^N(P_{r}(L_{k}) - Ave(P_{r}))^2,
\end{equation}

\noindent where $(\mathbb{X}, \mathbb{Y}, \mathbb{Z}, \mathbb{P})$ denotes the solution of objective function and it is expressed in detail in Eq. (\ref{solution}). ($\mathbb{X}, \mathbb{Y}, \mathbb{Z}$) represents the hovering positions of the UAVs for providing communication for distinct receivers. Note that, $\mathbb{Z}$ keeps constant for each UAV. Moreover, $\mathbb{P}$ is the transmission power of UAVs.

\begin{figure*}[!t]
	\normalsize
	\setcounter{MaxMatrixCols}{20}
	\begin{equation} 
		\label{solution}
		X=
		 [\mathbb{X}^{1 \times \mathcal{U}}, \mathbb{Y}^{1 \times \mathcal{U}}, \mathbb{Z}^{1 \times \mathcal{U}}, \mathbb{P}^{1 \times \mathcal{U}}]=[x^{1} ... x^{U}, y^{1} ... y^{U}, z^{1} ... z^{U}, P_{1} ... P_{U}]
	\end{equation}
	\hrulefill
	\vspace*{4pt}
\end{figure*}

\textbf{\emph{2) Optimization objective 2}}: Since there is an eavesdropper in this system and will certainly receive information, it is very necessary to guarantee the security of the system. Thereby, we increase the security by allowing the eavesdropper to receive as little information amount as possible, and it is provided as follows:

\begin{equation} 
	\label{transmission rate}
	C_{i,E} = \frac{1}{2}\log_2 \left(1+\frac{e}{2 \pi} \left (\frac{ (P_i h_{i,E})^2}{\sum_{r = 1, r \ne i}^{U} (P_r h_{r,E})^2 +\sigma_w} \right ) \right),
\end{equation}	

\noindent where $\sum_{r = 1, r \ne i}^{U} P_r h_{r,E}$ denotes the interference provided by other UAVs except $i$. In light of this, the objective function can be designed as below:

\begin{equation}
	f_2(\mathbb{X}, \mathbb{Y}, \mathbb{Z}, \mathbb{P}) = \sum_{i=1}^U(C_{i,E}),
\end{equation}

\textbf{\emph{3) Optimization objective 3}}: It is a better solution for the formulated UAVDMOP that the UAVs should migrate to more advantageous areas for communicating with receiving surfaces for the purpose of addressing the first two objectives. However, this procedure will consume excessive motion energy of UAVs. The objective function is to reduce the amount of motion energy that UAVs use as they get closer to the ideal hovering places, and is stated as follows.

\begin{equation}
	f_3(\mathbb{X}, \mathbb{Y}, \mathbb{Z}, \mathbb{P}) = \sum_{i=1}^{N} (E_{m}),
\end{equation}

\noindent where $E_m$ denote the motion energy consumed by UAV $i$, i.e., $E_m = P(V)t$. The time in the horizontal direction is denoted by $t$ and can be written as $t={L}/{V}$, in which $L$ denotes the distance between the starting point and the optimal hovering point of any UAV. Accordingly, the UAVDMOP in the considered UAV-enabled VLC system can be formulated as follows:

\begin{subequations}
	\label{MOP}
	\begin{align}
		\min _{X} & F=\left\{f_{1}, f_{2}, f_{3}\right\} \\
		\text { s.t. } & C1: X_{\min} \leqslant x_{i}^{U} \leqslant X_{\max}, \forall i, \\
		& C2: Y_{\min} \leqslant y_{i}^{U} \leqslant Y_{\max}, \forall i, \\
		& C3: h_{i}^U = H, \\
		& C4: P_{\min} \leqslant P_{i} \leqslant P_{\max}, \forall i, \\
		& C5: \omega_{i} \in \{0,1\},
	\end{align}
\end{subequations}

\noindent where constraints $C1$ and $C2$ jointly determine the bounds of horizontal search of UAVs. Constraint $C3$ represents the same altitudes for all UAVs. $P_{\min}$ and $P_{\max}$ denote the ranges of transmission power of UAVs. The last constraint $C5$ means that the regulator factors of UAVs are bounded in 0 $\sim$ 1.

%
%
\section{The proposed MOEA/D-CICM}
\label{The proposed MOEA/D-CICM}

\par For the purpose of handling the UAVDMOP formulated in the UAV-enabled VLC, a MOEA/D-CICM with two improved factors is proposed in this section.

%
%
\subsection{Motivation}
\par The single-objective optimization that considering only energy consumption or security will not meet the requirements of wireless network. However, MOPs are mainly used to solve the applications that need to optimize multiple conflicting objective functions at the same time. When solving the continuous MOPs, especially when it is difficult to assume the convexity or concavity of the problem, evolutionary algorithm (EA) is considered as a potential solution,~\citep{li2008multiobjective},~\citep{trivedi2016survey}. In recent years, numerous varieties of multi-objective evolutionary algorithms (MOEAs) have been put out, e.g., in~\citep{zhang2007moea},~\citep{tang2018adaptive} and~\citep{deb2002fast}. Among them, one of the algorithms that is most frequently used is MOEA/D\citep{zhang2007moea}, and it is generally regarded as a practical algorithm for solving MOPs considering that it exhibits the characteristics of quick convergence, good variety, and minimal computational complexity~\citep{wang2020improved},~\citep{liu2022multi}. The primary idea behind it is to break down the MOP into a group of single objective subproblems and solve them by applying population-based evolutionary method in a collaborative way. Moreover, it has been widely applied for solving a number of MOP applications, including multi-source scheduling in cloud computing~\citep{lin2017multi},~\citep{lin2017design} and parameter tuning in pattern recognition~\citep{li2008multiobjective},~\citep{zhou2019novel},~\citep{huang2018efficient}. To this end, we prefer to use it as the foundation algorithm in this article and improve it to make it better suited for dealing with the formulated UAVDMOP. Then, the following provides details on both the regular MOEA/D and the proposed MOEA/D-CICM.

%
%
\subsection{Conventional MOEA/D}

\par In the first part of this section, we thoroughly describe the traditional MOEA/D framework. By applying a decomposition approach in MOEA/D, which is usually Tchebycheff approach, an MOP can be broken down into $N$ separate scalar subproblems. The following expression can therefore be used to describe the objective function of the $i$th subproblem~\citep{zhang2007moea}:

\begin{equation}
	\label{function}
	g^{tc}\left(x^j \mid \chi^{j}, Z^*\right)=\max _{1 \leq i \leq t}\left\{\chi_{i}^{j}\left|f_{i}(x^j) - z_{i}^* \right|\right\},
\end{equation}

\noindent where for all $i \in \{1, ..., T \}$, $T$ represents the amount of objectives and $f_i(x)$ denotes the $i$th objective function. $x^j$ is the $j$th solution, where $j \in \{1, ..., N\}$ and $N$ are the index and the amount of solutions, respectively. $\chi_k^j = (\chi_1^j, \chi_2^j, ..., \chi_T^j)$ is $T$ evenly spread weight vectors and each $\chi_k^j$ satisfies $\sum_{k=1}^T{\chi_k^j}=1$ and $\chi_k^j \geq 0$ for all $k$ and $T$. Moreover, $Z^* = \{ z_1^*, z_2^*, ...,z_m^* \}$ is a group of reference vectors, and $z_i^* = \min \{f_i(x)|x \in \Omega \}$, which will be substituted by the minimum $f_i$ value found in the search. In search, MOEA/D holds~\citep{zhao2012decomposition}:

\begin{itemize}
	\item 1) A population of $N$ individuals $x_1, x_2, ..., x_N \in \Omega$, $x_j$ is the current solution to $j$th subproblem;
	\item 2) ${FV}_1, ..., {FV}_N$, where ${FV}_j$ is $F$-value of $x_j$, i.e., ${FV}_j = F(x_j)$ for each $j = 1, ..., N$;
	\item 3) $Z^* = (z_1^*, ..., z_t^*)$, where $z_i^*$ is the current best value for the objective function value $f_i$.
\end{itemize}

\par Moreover, in conventional MOEA/D, each solution and each subproblem have its own neighborhood size (NS)-neighborhood, respectively, which is the collection of weight vectors from NS-closet for each weight vector. Moreover, similar to most other MOEAs, MOEA/D compares different solutions also based on the pareto optimal dominance in solving MOPs. In addition, the weight vectors of MOEA/D are mainly scattered near the edge of the weight space, which causes the optimal solutions to be spread out near the edge of PF~\citep{zheng2018improved}. Algorithm \ref{Algorithm: 1} provides the primary steps of conventional MOEA/D.

\setlength{\textfloatsep}{0.1cm}
\begin{algorithm}[h!]
	\caption{Conventional MOEA/D.}\label{Algorithm: 1}
	\quad 1) We respectively set the amount of receivers and UAVs to $N$ and $U$, the maximum number of iterations to ${It}_{max}$, the size of population to $Q$. Initialize randomly the ideal point $z^*$, where $z^* = \min(f_i(x)|x \in \Omega)$, and the locations of UAVs and receivers; Initialize the population ${X}^1$, ..., ${X}^Q$; \\
	\quad 2) Use the uniform design strategy to generate a large number of weight vectors $\chi_{1}, ..., \chi_{T}$; \\
	\quad 3) The Euclidean distances between each $chi_i (i=1,..., T)$ and other weight vectors are calculated; \\
	\quad 4) The mating neighborhood size $M_t$ and the replacement neighborhood size $M_r$ are generated at random; \\
	\quad 5) Evaluate the fitness function value $f_{X_i}$ of each solution ${X}_i$ in the population; \\
	\For{$it=1$ to $It_{max}$}{
		\For{$i=1$ to $U$} {
			\quad By adopting the crossover operation of a genetic algorithm, two solutions (i.e., $X^i_{1}$ and $X^i_{2}$) are randomly chosen from the mating neighborhood $M_t$ of population $X^i$, and used to build the novel solution $X_{new}$; \\
			\quad Evaluate the fitness function $f_{X_{new}}$ of the new solution $X_{new}$; \\
			\quad $z^*= \min \{z^*, f_{X_{new}}\}$ is Updated; \\ 
			\For{$X_j$ in the replacement neighborhood $M_r$ of $X_i$}
			{\If{$g^{(tc)}(X_{new}|\chi^{j}, z^*) \le g^{(tc)}(X_{j}|\chi^{j},z^*)$} {$X_j = X_{new}$ and $f_{X_j} = f_{X_{new}}$}}
		} 
	} 
\end{algorithm}
\setlength{\textfloatsep}{0.1cm}

%
%
\subsection{MOEA/D-CICM}

\par For the sake of solving the formulated UAVDMOP in UAV-enabled VLC, MOEA/D-CICM is proposed in this section to improve the drawbacks of standardized MOEA/D. Specifically, to strengthen the performance of MOEA/D, two improved factors are included, which are the chaos initialization and crossover mutation strategies. The details of the overall MOEA/D-CICM framework are presented in Algorithm~\ref{Algorithm: 2}:

\begin{algorithm}[h!]
	\caption{MOEA/D-CICM.}
	\label{Algorithm: 2}
	Initialize the transmission power of UAVs according to the chaos initialization operator, i.e., Eq. (\ref{circle map}). The remaining parameters and variables are initialized according to Algorithm \ref{Algorithm: 1}; \\
	\For{$it=1$ to $It_{max}$}{
		\For{$i=1$ to $U$} {
			\quad Update the solutions with different physical meanings by using Algorithm \ref{Crossover mutation strategy}, and combine them into a novel solution $X_{new1}$, then judge and limit its boundary, and obtain the individual $X_{new}$; \\
			\quad Evaluate the fitness function  $f_{X_{new}}$ of the new individual $X_{new}$; \\
			\quad Update $z^*= \min \{z^*, f_{X_{new}}\}$; \\ 
			\For{$X_j$ in the replacement neighborhood $M_r$ of $X_i$}
			{\If{$g^{(tc)}(X_{new}|\chi^{j}, z^*) \le g^{(tc)}(X_{j}|\chi^{j}, z^*)$} {$X_j = X_{new}$ and $f_{X_j} = f_{X_{new}}$;} }
		} 
	} 
\end{algorithm}

%
%
\subsubsection{Chaos initialization strategy}

\par As can be seen in most MOEAs, including MOEA/D, the population is randomly constructed. The range of potential solutions, as well as their diversity and quality, will be constrained by the random initialization method. Moreover, the search step size of the solutions is not easy to be controlled. In addition, the experiment results also demonstrate that in the formulated UAVDMOP, the variation range of solutions is relatively tiny or almost constant as a result of random initialization, which makes it easy for the solutions to enter the local optimal in subsequent iterations. Thereby, the initialization operator of the population has to be improved urgently.

\par The chaos theory is proposed by Poincare, and later extended by Lornez~\citep{lorenz1963deterministic} to solve the unpredictable complex nonlinear problems~\citep{dhanya2010nonlinear}. A chaotic system can be defined by its trajectory in state space and is predictable and dynamic, evolving from the given initial circumstances~\citep{sarveswararao2022optimal}. Moreover, in order to spread the solutions more uniformly across the search space, chaos theory offers the capacity to transfer optimization variables from the chaotic space domain to solution space. It is shown that the random search is inferior than optimization search with chaotic variables~\citep{zhang2009application}. Since the solutions of the formulated UAVDMOP include the transmission power and hovering positions of UAVs, whose effective ranges are distinctive, we use different methods to initialize them separately to improve the quality and search space of each solution. Moreover, the altitude of UAVs is fixed in this work, thus, only the horizontal positions and transmission power of UAVs need to be optimized. Therefore, the circle map is used in the initialization of transmission power of UAVs since it improves the diversity of the solutions and avoids the solutions falling into local optimal with respect to the formulated UAVDMOP, while the horizontal locations of UAVs are still random initialized. Accordingly, the circle map are given as follows:
\begin{equation} \label{circle map}
	\mathbb{X}_{it} =\mod(\mathbb{X}_{it} + d-(c/(2 * \pi)) * \sin(2 * \pi * \mathbb{X}_{it}), 1),
\end{equation}

\noindent where $\mod$ represents the mathematical modular division operation, and $it$ stands for the index of the solution dimension in chaotic sequence. $c$ and $d$ are two constants~\citep{saremi2014biogeography}. $\mathbb{X}_{it}$ is the value of the $it$th dimension with respect to the transmission power of UAVs.

%
%
\subsubsection{Crossover mutation strategy}

\par In conventional MOEA/D, the replication operation mainly adopts the crossover operation of genetic algorithm to produce offspring individuals. This operation limits the capacities of exploration and exploitation of population in the formulated UAVDMOP. Therefore, in order to improve the efficiency of traditional MOEA/D, we employ the uniform crossover and mutation operator of differential evolution (DE) in this section. Each part of the solutions needs to be improved separately in this work since the solutions of the formulated UAVDMOP consist of different components with different physical meanings, and their lower and upper boundaries are different. To this end, each solution firstly is divided into three parts, that are the horizontal locations, vertical positions, and transmission power of UAVs. Then, in terms of the power and horizontal locations of UAVs, uniform crossover and mutation operation of DE is used to improve it by limiting distinct boundaries for them. The specific operation process of solution update is given in Algorithm \ref{Crossover mutation strategy}, and it is described in detail as follows:

\begin{algorithm}[h!]
	\caption{Crossover mutation strategy.}
	\label{Crossover mutation strategy}
	From the mating neighborhood $M_m$ of population $X^i$, Select two individuals $X^i_{1}$ and $X^i_{2}$; \\
	Based on the distinct physical meanings, the two chosen individuals are separated into ${X^i_{11}}$, ${Z^i_{11}}$, ${P^i_{11}}$ and ${X^i_{12}}$, ${Z^i_{12}}$, ${P^i_{12}}$; \\
	Crossover the two new solutions ${X^i_{11}}$ and ${X^i_{12}}$ and obtain ${X_{new}^{1}}$ and ${X_{new}^{2}}$ by using Eq. (\ref{uniform crossover}).
	$Z$ is the altitude of the UAVs and remains unchanged, and the operation of $P$ is consistent with that of $X$. Determine and constrain the boundaries of the new solution components $X_{new}$ and $P_{new}$; \\
	Mutate the solutions after crossover according to Eq. (\ref{mutation}). \\
	Combine each component of the solutions after crossover and mutation to form the final solution of the each individual.
\end{algorithm}

\textbf{For the uniform crossover}, similar to~\citep{syswerda1989uniform}, the proposed uniform crossover refers to setting a specific value to determine whether the corresponding genes are inherited from the parent individuals to the new individuals or not. The operation process is as follows: when the gene in individual $A$ is less than or equal to the specified value, the relevant genes from the previous individual $A$ are still passed on to the new individual $A$. While the new individual $A$ inherits the appropriate gene from the other old individual $B$ when the gene is greater than the given value. The above operators generate a fully new individual $A'$. Undoubtedly, the uniform crossover is a part of the multi-point crossover and can be stated as below.

\begin{equation}
	\label{uniform crossover}
	\mathbb{X}_{A^{'}}(i) = 
	\begin{cases}
		\mathbb{X}_A(i), & \mathbb{X}_A(i) \leq {X}, \\ 
		\mathbb{X}_B(i), & \rm{otherwise},
	\end{cases}
\end{equation}

\noindent where ${X}$ is the value given according to different solution ranges, which is different with respect to different physical meaning. Consequently, it is set as half of the valid value range in this work. In addition, both $\mathbb{X}_A$ and $\mathbb{X}_B$ are the solution components of two different individuals with the same physical meanings.

\begin{figure*}
	\centering
	\includegraphics[width=5.5in]{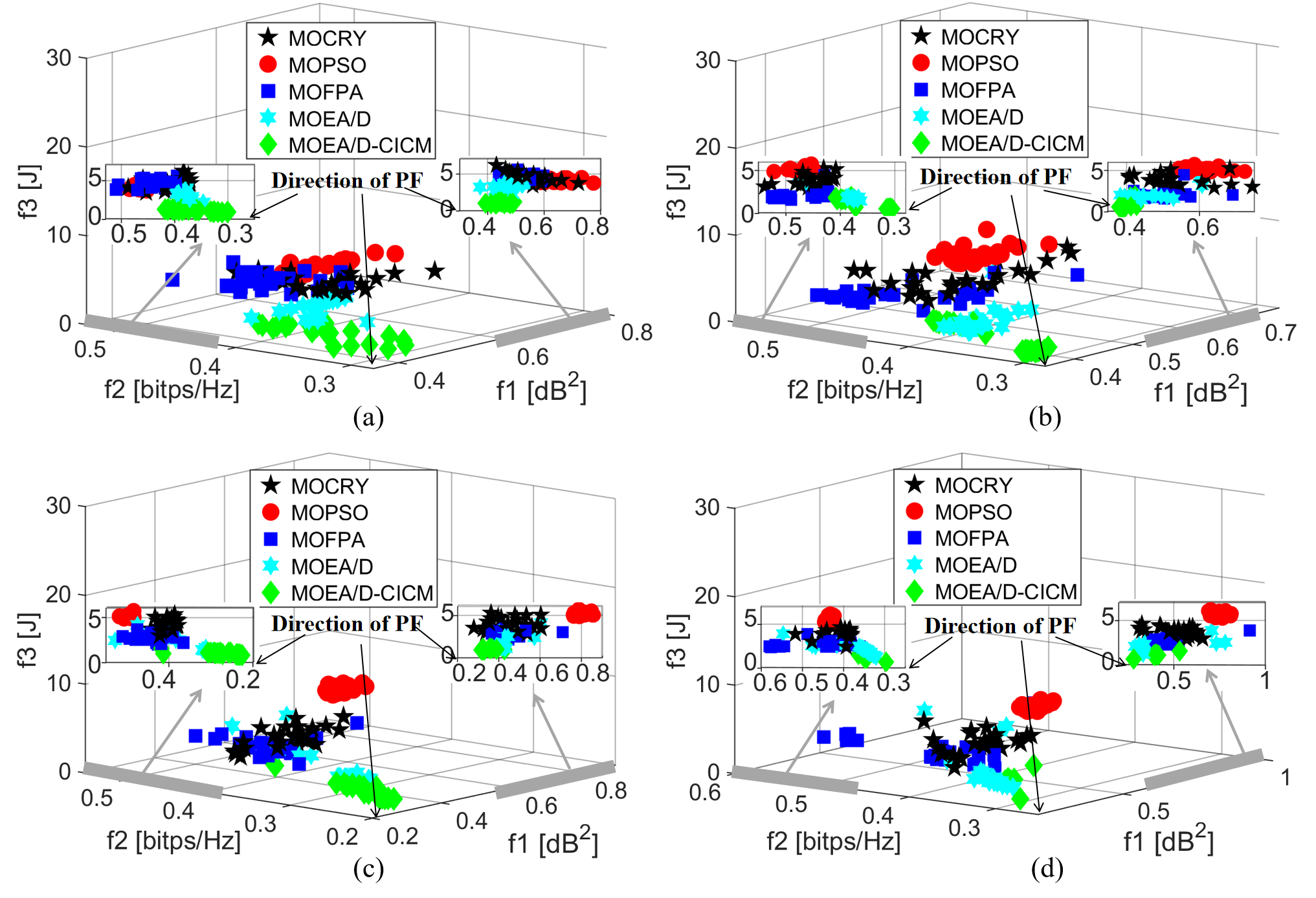}
	\caption {Solution distributions achieved by different algorithms in Case 1 (8 UAVs) at different iterations. (a) 50th iteration. (b) 100th iteration. (c) 150th iteration. (d) 200th iteration.}
	\label{Solution distributions achieved by different algorithms in Case 1 (8 UAVs) at different iterations}
\end{figure*}

\textbf{From the perspectives of mutation operator}, DE algorithm is a new parallel direct search EA for solving continuous optimization problems~\citep{vesterstrom2004comparative}. The mutation operation of the conventional DE is to add a difference vector to base vector, and the difference vector controls the shift direction of the base vector. Although the mutation in conventional DE can enhance the exploration ability of solutions, the base vector and difference vector are randomly selected, which will lead to inefficient solutions. Thus, we propose a new mutation operator to enhance the searching efficiency and quality of solutions, and it can be designed as follows:

\begin{equation}
	\label{mutation}
	\mathbb{X}_{A^{'}} = \mathbb{X}_{B^{'} }- (\mathbb{X}_{B^{'}} - \mathbb{X}_{A^{'}})*rand* \beta ,
\end{equation}

\noindent where $\mathbb{X}_{A^{'}}$ and $\mathbb{X}_{B^{'}}$ represent two different solutions with the same physical meaning obtained from the uniform crossover operator. $rand$ is a random number and $\beta$ can be calculated according to Eq. (\ref{beta}).

\begin{equation}
	\label{beta}
	\beta = 
	\begin{cases}
		(2Q)^{\frac{1}{n+1}}-(it-\frac{it}{\sqrt{{It}_{max}}})/(\sqrt{{It}_{max}*n}),& Q < P_c, \\
		\frac{1}{2}(1-Q)^{\frac{1}{n+2}} + (it-\frac{it}{\sqrt{{It}_{max}}})/(\sqrt{{It}_{max}*n}),& Q \ge P_c.
	\end{cases}
\end{equation}

\noindent where $Q$ and $P_c$ are random number and mutation probability, respectively. $n$ is a constant and is set to be 3 in our simulations. $it$ and ${It}_{max}$ denote the current number of iterations and the total number of iterations, respectively. Finally, after crossover and mutation, each component of the solutions is integrated to create the final solution of each individual. The pseudocode of crossover mutation strategy is described in Algorithm \ref{Crossover mutation strategy}.

%
%
\subsection{Complexity of the proposed algorithm}

\par The random initialization process used by other methods is still maintained in this study. The complexity of the proposed approach remains same despite the fact that the chaotic initialization operator enhances the performance of initialization solutions, because there is only one inner loop in these initialization operations. The reproduction methods of solution mainly account for the differences in algorithm complexity, thereby, we need to elaborate it in detail.

\par It is assumed that $G$ represents the maximum number of iterations. The quantities of receivers and UAVs are respectively set to $N$ and $U$, and $N \gg U$ in this work. There is just one inner loop, hence the complexity is $O(N)$ when distinct solutions are divided into multiple parts according to their diverse physical meanings. The update procedure of solutions with crossover mutation operator is the same as that of division operator, which means that there is only one inner loop in this operator, thereupon the complexities is also ${O(N)}$. In conclusion, the overall complexity of MOEA/D-CICM is ${O(G N U)}$.

\begin{figure*}
	\centering
	\includegraphics[width=6in]{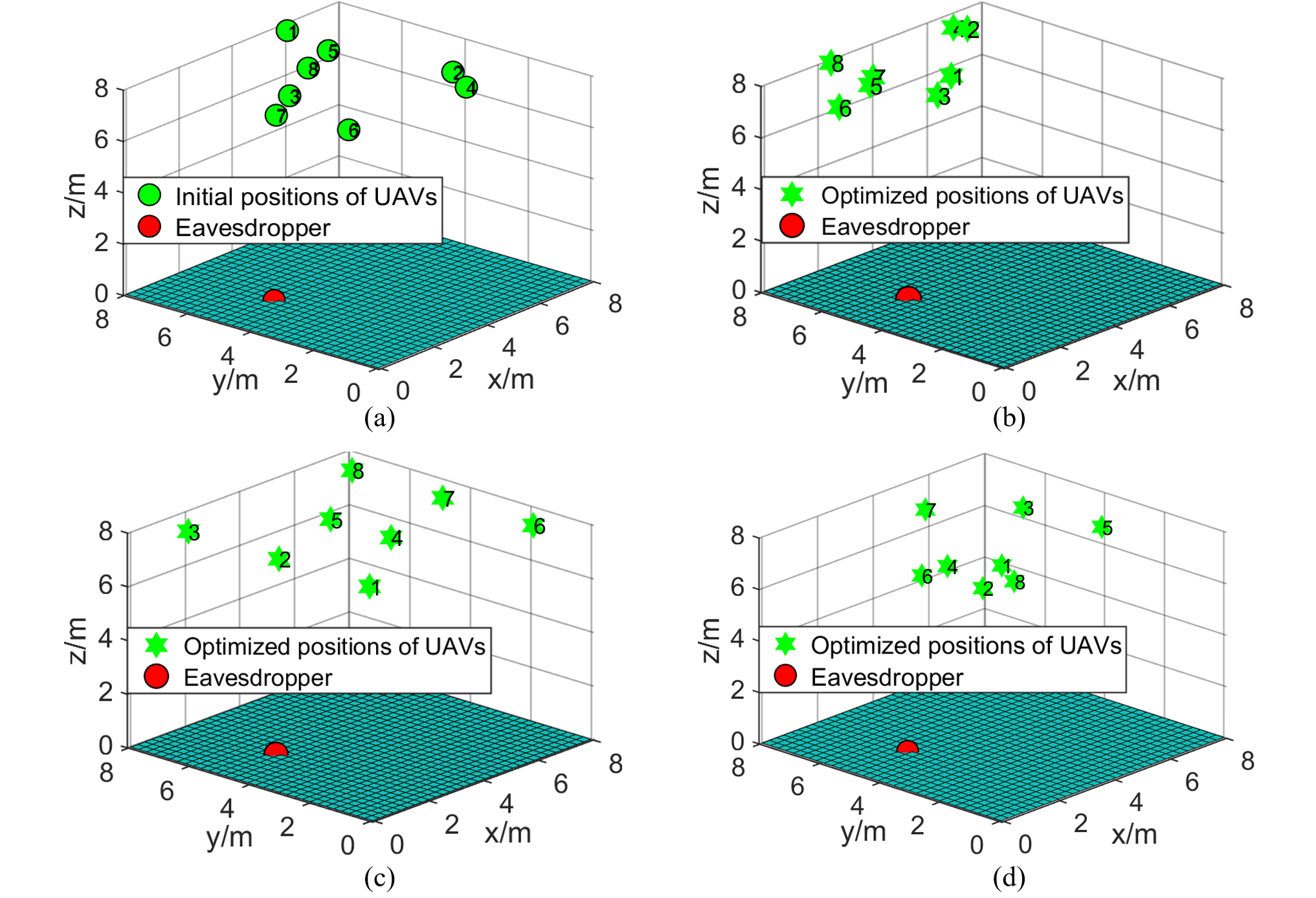}
	\caption {The deployments of UAVs obtained by different methods for Case 1 (8 UAVs). (a) Initial deployments of UAVs. (b) Random deployments. (c) Uniform deployment. (d) MOEA/D-CICM.}
	\label{Deployment of 8 UAVs.}
\end{figure*}

%
%
\section{Simulation results}
\label{Simulation results}

%
%
\subsection{Simulation setups}

\par So as to evaluate the performance of the proposed MOEA/D-CICM in solving the formulated UAVDMOP of UAV-enabled VLC system, we conduct simulations in Matlab 2021a. It is assumed that the power range of UAVs ($P_{\min}$ and $P_{\max}$) is set to 0.1 $\sim$ 10 W, respectively. In addition, $V$ is set to 16 m/s, and Table \ref{System parameters} lists the remaining parameters. For the setups of the algorithm, in order to prevent random bias, we execute each method 10 times with a maximum of 200 iterations.

\begin{table}[htbp]
	\centering
	\caption{System parameters}
	\begin{tabular}{c|c} \hline
		Parameters & Values \\ \hline
		Semi-angle at half power $\Phi_{1/2}$ & {$60^{\circ}$} \\ \hline
		Receiver semi-angle of FOV $\Psi_{c}$ & {$60^{\circ}$} \\ \hline
		Detect area of photodiodes $A_r$ & 1 {cm$^2$} \\ \hline
		Refractive index $n_r$ & 1.5 \\ \hline
		Power adjustment factor $\omega_{i}$ & {0 $\sim$ 1} \\\hline
		Rotor disc area $A$ & 0.503 \\\hline
		Mean rotor induced velocity during hovering $v_0$ & 4.03 \\\hline
		The fuselage drag ratio $d_0$ & 0.6 \\\hline
		Tip speed of the rotor blade $U_{tip}$ & 120 \\\hline
		Rotor solidity $s$ & 0.05 \\\hline
		Air density $\rho$ & 1.225 \\\hline
		Standard deviation of the AWGN $\sigma_w$ & -110 dB \\\hline
	\end{tabular}
	\label{System parameters}
\end{table}

\par We conduct two categories of simulations in this work. On the one hand, in terms of the UAV deployment methods, we add random deployment and uniform deployment methods. The UAVs are randomly distributed in monitoring region when using random methods, and when using uniform deployment methods, they are distributed uniformly at a predetermined altitude. On the other hand, for the reason that to confirm the effectiveness of the proposed algorithm from an algorithmic perspective, several comparison approaches are adopted, including MOCRY~\citep{khodadadi2021multi}, MOPSO~\citep{coello2002mopso}, MOFPA~\citep{yang2014flower}, in addition to the conventional MOEA/D. Note that the proposed updating approach of dividing the solutions into several parts is adopted in these comparison algorithms, which makes them have the ability to deal with the components with different lower and upper boundaries in this work.

%
%
\subsection{Optimization results}

\begin{figure*}
	\centering
	\includegraphics[width=6.2in]{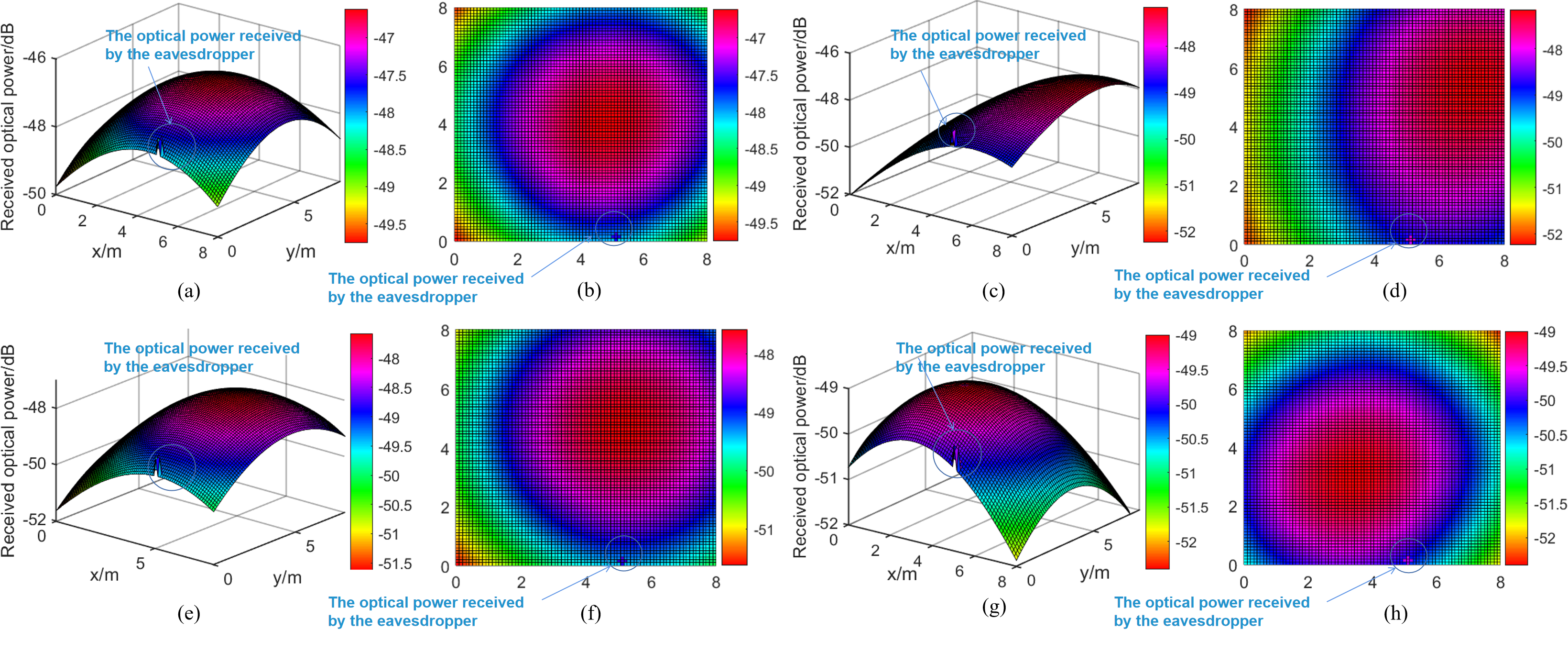}
	\caption {The maps of the received optical power distribution obtained by different methods for Case 1 (8 UAVs). (a) 3D power distributions at initial deployment of UAVs. (b) 2D power distributions at initial deployment of UAVs. (c) 3D power distributions of random deployment. (d) 2D power distributions of random deployment. (e) 3D power distributions of uniform deployment. (f) 2D power distributions of uniform deployment. (g) 3D power distributions of MOEA/D-CICM. (h) 2D power distributions of MOEA/D-CICM.}
	\label{Optical power of 8 UAVs.}
\end{figure*}

\par In this section, the formulated UAVDMOP is solved via using the proposed MOEA/D-CICM and other abovementioned comparison algorithms, and compare the outcomes to the random and uniform deployment methods. In particular, we carry on two situations of simulations, where the monitor area is set to 8 m $\times$ 8 m, and the number of UAVs and receivers is set to be 8 and 6400, respectively. Second, the monitor area is set to 10 m $\times$ 10 m, and the quantities of receivers and UAVs are set to be 10000 and 12, respectively, and the altitudes of UAVs in these two cases are fixed to 8 m.

\begin{figure*}
	\centering
	\includegraphics[width=5.5in]{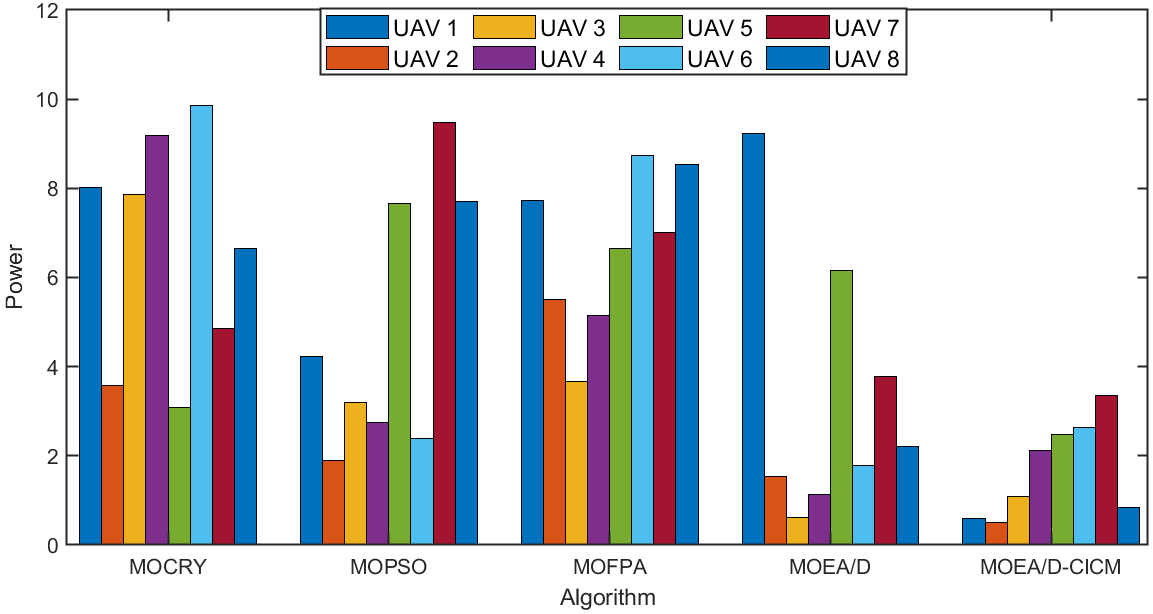}
	\caption {Transmission power of 8 UAVs obtained by different algorithms.}
	\label{Transmission power of 8 UAVs obtained by different algorithms.}
\end{figure*}

%
%
\subsubsection{Cases 1-Optimization results of 8 UAVs}

\par Table \ref{Numerical results obtained by different methods for case 1 (8 UAVs)} shows the numerical results for variance $f_{1}$, the amount of information the eavesdropper obtained $f_{2}$, and the total motion energy consumed by UAVs $f_{3}$. It can be seen from this table, when compared to other methods, the proposed MOEA/D-CICM performs the best in terms of these optimization objectives.

\begin{table}[htb]
	\centering
	\caption{Numerical results obtained by different methods for case 1 (8 UAVs)}
	{\begin{tabular}{|c|c|c|c|c|} \hline
			{\bfseries Algorithm} & {$f_1$({\bfseries dB$^2$})} & {$f_2$({\bfseries bps/Hz})} & {$f_3$({\bfseries J})} \\ \hline
			Random	&1.7011 &	0.6173 & 4.5092 \\\hline
			Uniform & 0.6786 & 0.5649 & 3.6322 \\\hline
			MOCRY	& 0.4853 & 0.4135 & 3.6831 \\\hline
			MOPSO & 0.7469 & 0.4316 & 5.5772 \\\hline
			MOFPA & 0.4566 & 0.4615 & 2.5855 \\\hline
			MOEA/D & 0.3709 & 0.3634 & 1.9556 \\\hline
			\bfseries{MOEA/D-CICM} & \bfseries{0.3176} & \bfseries{0.312} & \bfseries{0.7138} \\\hline
		\end{tabular}
		\label{Numerical results obtained by different methods for case 1 (8 UAVs)}}
\end{table}

\begin{figure*}
	\centering
	\includegraphics[width=5.5in]{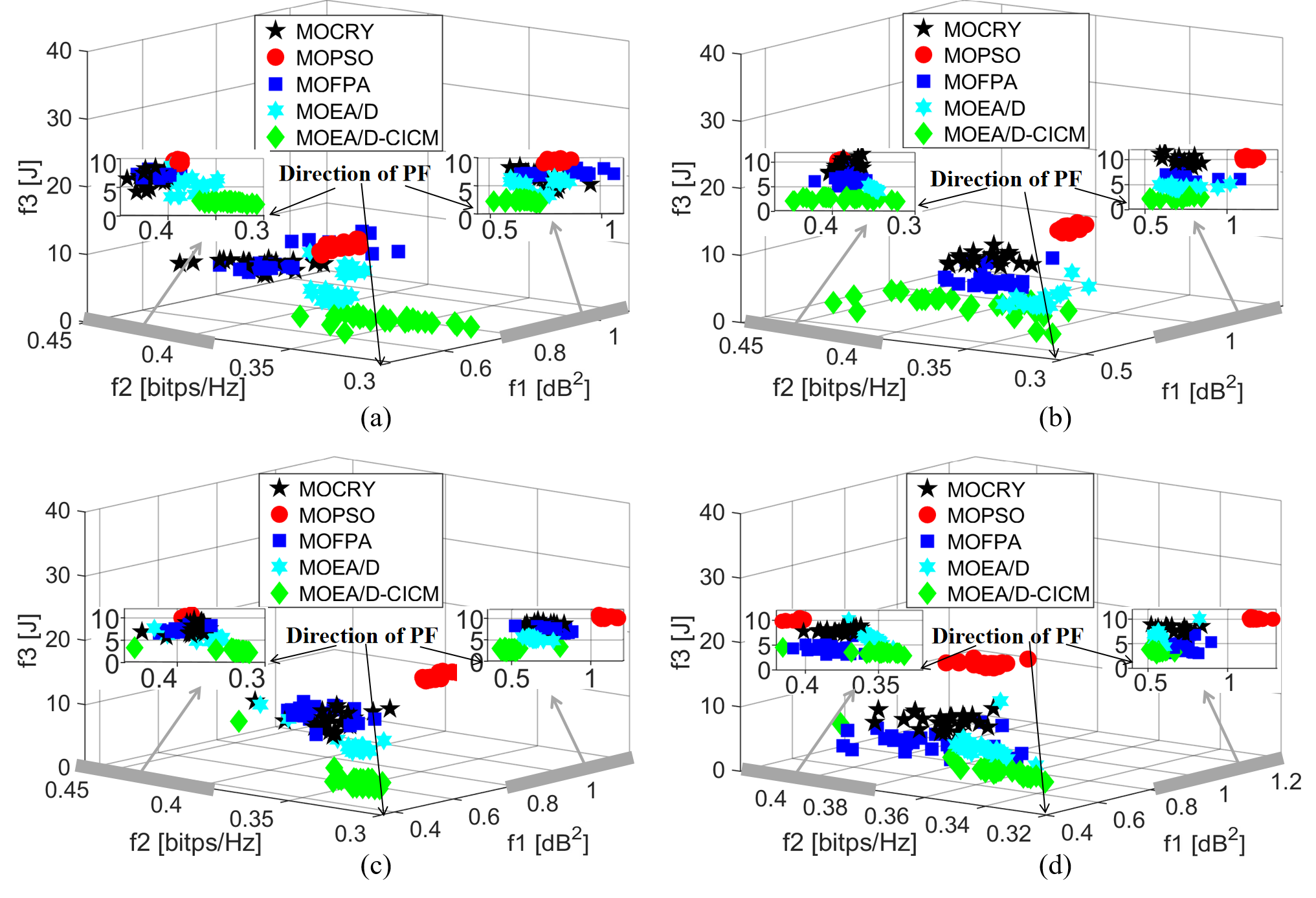}
	\caption {Solution distributions achieved by different algorithms in Case 2 (12 UAVs) at different iterations. (a) 50th iteration. (b) 100th iteration. (c) 150th iteration. (d) 200th iteration.}
	\label{Solution distributions achieved by different algorithms in Case 2 (12 UAVs) at different iterations}
\end{figure*}

\begin{figure*}
	\centering
	\includegraphics[width=6in]{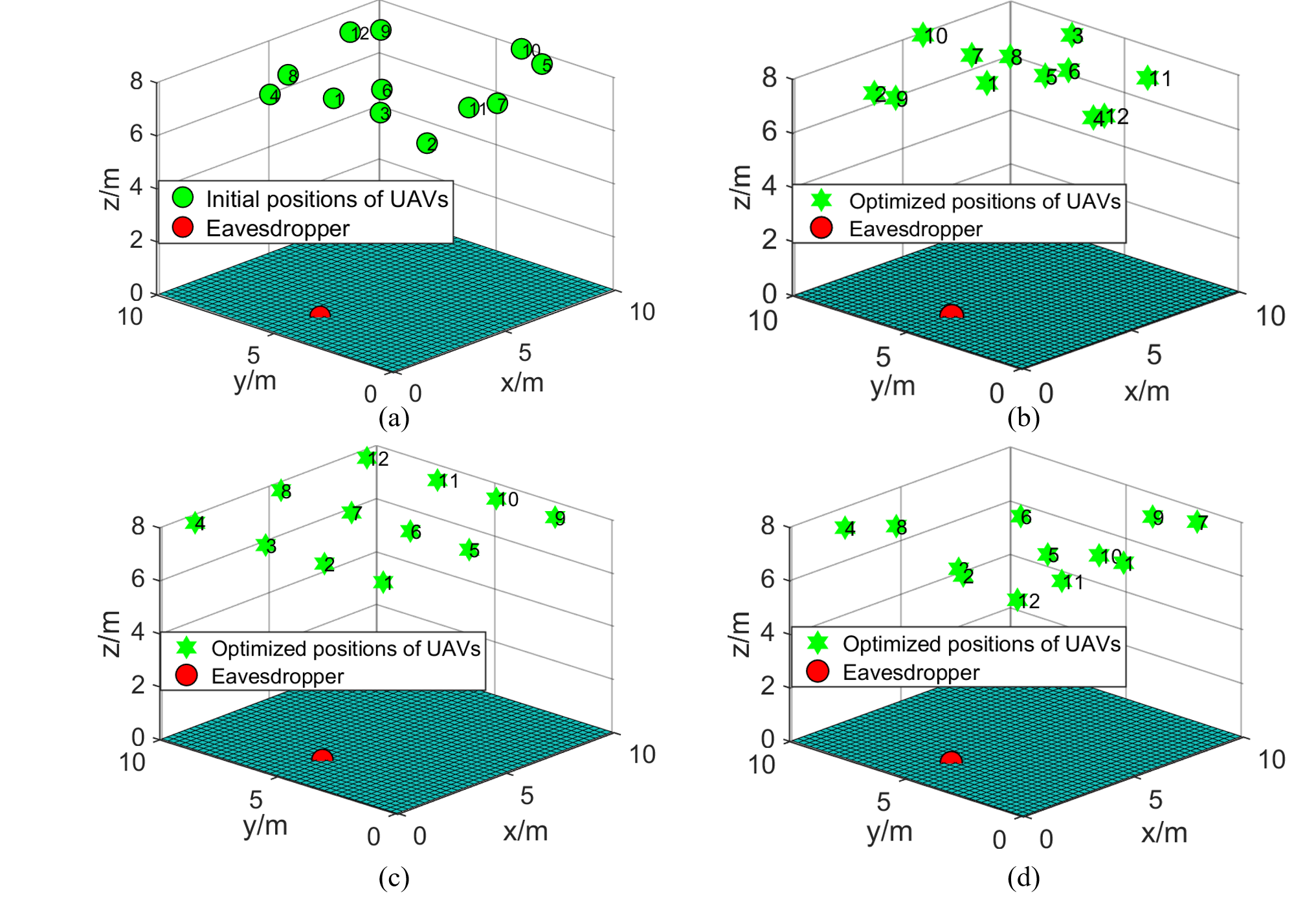}
	\caption {The deployments of UAVs obtained by different methods for Case 2 (12 UAVs). (a) Initial deployments of UAVs. (b) Random deployments. (c) Uniform deployment. (d) MOEA/D-CICM.}
	\label{Deployment of 12 UAVs.}
\end{figure*}

\par The solution distributions obtained by several MOEAs at various iterations are provided in Fig. \ref{Solution distributions achieved by different algorithms in Case 1 (8 UAVs) at different iterations} in a more understandable manner. As we can see from these figures that, when the number of iterations is increased, solutions achieved by employing the proposed MOEA/D-CICM are more near to the ideal position (i.e., the PF depicted in these figures) than those of other methods. That is to say, for addressing the formulated UAVDMOP, the proposed MOEA/D-CICM is more superior. It is noted that these values are the average over various iterations. 

\begin{figure*}
	\centering
	\includegraphics[width=6.2in]{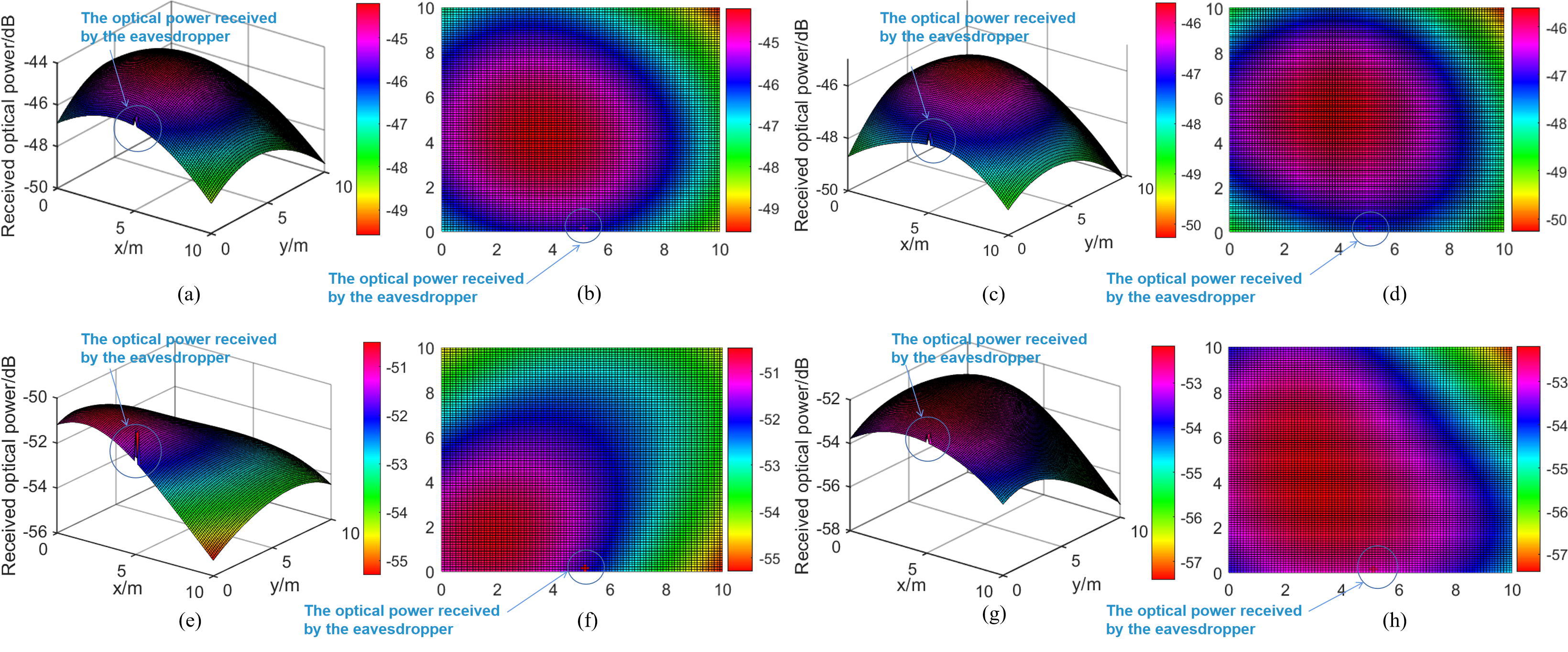}
	\caption {The maps of the received optical power distribution obtained by different methods for Case 2 (12 UAVs). (a) 3D power distributions at initial deployment of UAVs. (b) 2D power distributions at initial deployment of UAVs. (c) 3D power distributions of random deployment. (d) 2D power distributions of random deployment. (e) 3D power distributions of uniform deployment. (f) 2D power distributions of uniform deployment. (g) 3D power distributions of MOEA/D-CICM. (h) 2D power distributions of MOEA/D-CICM.}
	\label{Optical power of 12 UAVs.}
\end{figure*}

\par The deployment of 8 UAVs, which was accomplished using various techniques, is shown in Fig. \ref{Deployment of 8 UAVs.}. It should be noted that, the solid green circles with Arabic numbers in Fig. \ref{Deployment of 8 UAVs.}(a) indicate the starting location of the $i$th UAV, and the solid green stars with numbers in Figs. \ref{Deployment of 8 UAVs.}(b), \ref{Deployment of 8 UAVs.}(c), and \ref{Deployment of 8 UAVs.}(d) indicate the optimal locations of these UAVs, which are obtained by using random deployment method, uniform deployment method and the proposed MOEA/D-CICM, respectively.

\par Moreover, maps of the received optical power distributions for 8 UAVs are shown in Fig. \ref{Optical power of 8 UAVs.}, which are obtained by utilizing various deployment techniques. Especially, Figs. \ref{Optical power of 8 UAVs.}(a) and \ref{Optical power of 8 UAVs.}(b) respectively show the 3D and 2D maps of the optical power distribution that is received when UAVs are deployed in their initial positions. Similarly, in Figs. \ref{Optical power of 8 UAVs.}(c), \ref{Optical power of 8 UAVs.}(e), \ref{Optical power of 8 UAVs.}(g), and \ref{Optical power of 8 UAVs.}(d), \ref{Optical power of 8 UAVs.}(f) and \ref{Optical power of 8 UAVs.}(h), which are achieved via employing other methods, respectively are 3D and 2D maps of the received optical power distribution. The acreage within the blue zone (i.e., the high quality communication area) in Figs. \ref{Optical power of 8 UAVs.}(b) and \ref{Optical power of 8 UAVs.}(d) are roughly 40.97 m{$^2$} and 43.34 m{$^2$}, while they are respectively roughly 44.58 m{$^2$} and 50.19 m{$^2$} in Figs. \ref{Optical power of 8 UAVs.}(f) and \ref{Optical power of 8 UAVs.}(h), and the high quality communication area of the received optical power is increased by 22.5$\%$, 15.81$\%$ and 12.58$\%$ after using the proposed MOEA/D-CICM. Moreover, Fig. \ref{Transmission power of 8 UAVs obtained by different algorithms.} illustrates the transmission power of 8 UAVs accomplished by various algorithms.

%
%
\subsubsection{Case 2-Optimization results of 12 UAVs}

\par The statistical results with respect to the three objective functions are presented in Table \ref{Numerical results obtained by different methods for case 2 (12 UAVs)}, from which we can see that the proposed MOEA/D-CICM achieves the highest values in terms of these objectives.

\begin{table}[htb]
	\centering
	\caption{Numerical results obtained by different methods for case 2 (12 UAVs)}
	{\begin{tabular}{|c|c|c|c|c|} \hline
			{\bfseries Algorithm} & {$f_1$({\bfseries dB$^2$})} & {$f_2$({\bfseries bps/Hz})} & {$f_3$({\bfseries J})} \\ \hline
			Random	&0.7856 &	0.4060 & 7.8843 \\\hline
			Uniform &1.3068 & 1.3557 & 1.7727 \\\hline
			MOCRY	& 0.6788 & 0.3779 & 7.9391 \\\hline
			MOPSO & 1.1716 & 0.4048 & 10.0484 \\\hline
			MOFPA & 0.6843 & 0.3842 & 4.2658 \\\hline
			MOEA/D & $0.5703$ & $0.3544$ & 6.3541 \\\hline
			\bfseries{MOEA/D-CICM} & \bfseries{0.5577} & \bfseries{0.3504} & \bfseries{3.5071} \\\hline
		\end{tabular}
		\label{Numerical results obtained by different methods for case 2 (12 UAVs)}}
\end{table}

\par Comparable to instance 1, Fig. \ref{Solution distributions achieved by different algorithms in Case 2 (12 UAVs) at different iterations} depicts the more understandable outcomes of the solution distributions attained by various methods throughout multiple iterations. The suggested MOEA/D-CICM performs better than other algorithms due to its proximity to the PF, as can be observed from the figures. In addition, the deployment of 12 UAVs obtained by various approaches is shown in Fig. \ref{Deployment of 12 UAVs.}, in which Figs. \ref{Deployment of 12 UAVs.}(a), \ref{Deployment of 12 UAVs.}(b), \ref{Deployment of 12 UAVs.}(c) and \ref{Deployment of 12 UAVs.}(d) represent the results of initial deployment, random deployment, uniform deployment, and the optimal deployment obtained by the proposed MOEA/D-CICM, respectively. Moreover, the maps of the received optical power distributions optimized by various methods for 12 UAVs are provided in Fig. \ref{Optical power of 12 UAVs.}, where Figs. \ref{Optical power of 12 UAVs.}(a), \ref{Optical power of 12 UAVs.}(c), \ref{Optical power of 12 UAVs.}(e) and \ref{Optical power of 12 UAVs.}(g) denote the 3D maps of the received optical power distribution achieved by the abovementioned various methods. Similarly, the 2D maps achieved by these approaches are give in Figs. \ref{Optical power of 12 UAVs.}(b), \ref{Optical power of 12 UAVs.}(d), \ref{Optical power of 12 UAVs.}(f) and \ref{Optical power of 12 UAVs.}(h), respectively. The acreage within the high quality communication area in Figs. \ref{Optical power of 12 UAVs.}(b) and \ref{Optical power of 12 UAVs.}(d) are about 62.11 m{$^2$} and 59.25 m{$^2$}, while they are respectively about 50.69 m{$^2$} and 63.33 m{$^2$} in Figs. \ref{Optical power of 12 UAVs.}(f) and \ref{Optical power of 12 UAVs.}(h), and the high quality communication area of the received optical power is increased by 1.96$\%$, 6.89$\%$ and 24.94$\%$ after using the proposed MOEA/D-CICM. These obtained findings indicate that the suggested MOEA/D-CICM outperforms others in tackling the formulated UAVDMOP.

\begin{figure*}
	\centering
	\includegraphics[width=6in]{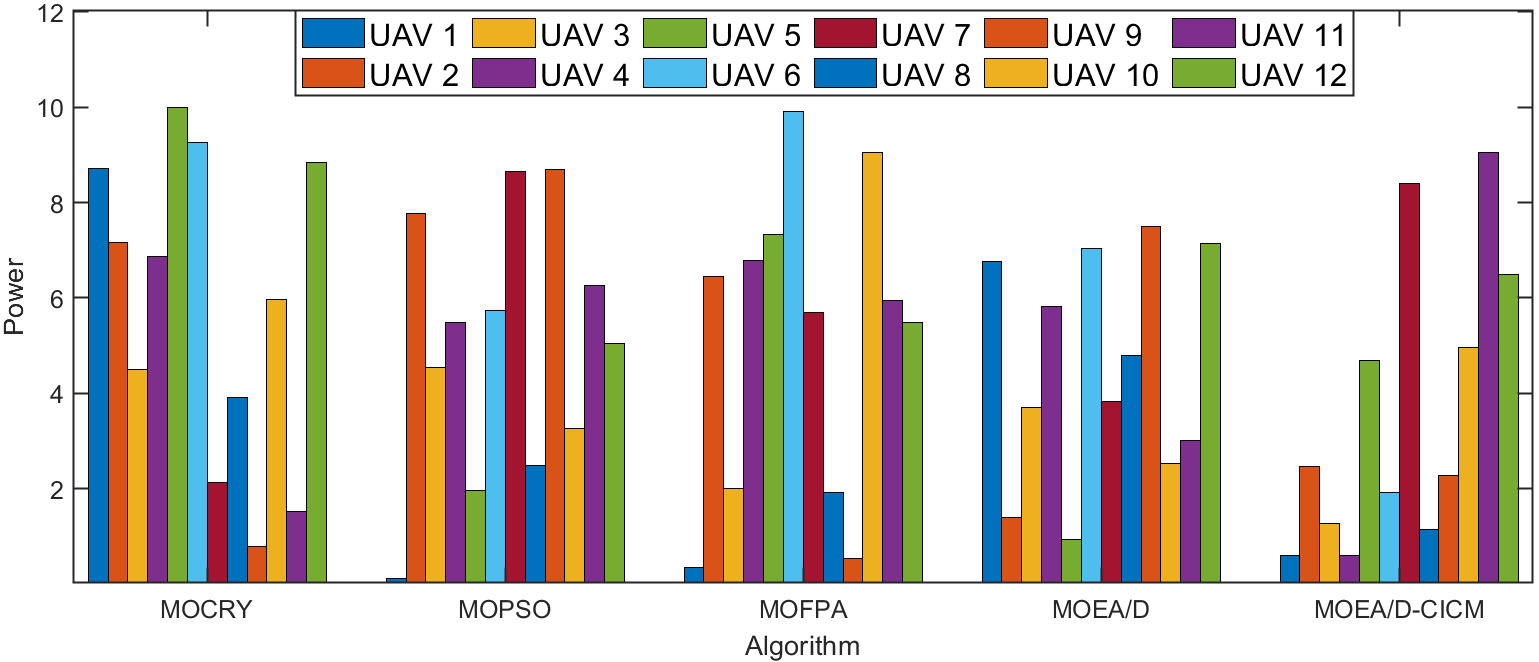}
	\caption {Transmission power of 12 UAVs obtained by different algorithms.}
	\label{Transmission power of 12 UAVs achieved by different algorithms.}
\end{figure*}

In addition, the transmission power attained by 12 UAVs using various MOEAs are indicated in Fig. \ref{Transmission power of 12 UAVs achieved by different algorithms.}.

%
%
\section{Conclusion}
\label{Conclusion}

\par In this work, the UAV-enabled VLC is analyzed, where numerous UAVs are employed for providing communication service for terrestrial receiving surface, in which an unauthorized eavesdropper is existed. First, a UAVDMOP is formulated to jointly make the optical power received by the receiving surface more uniform, minimize the amount of information captured by the eavesdropper, and minimize the total motion energy consumption of UAVs via cooperatively optimizing the locations and transmission power of UAVs. Then, so as to deal with this issue, a MOEA/D-CICM that incorporates three modified factors is proposed, and comprehensive simulation outcomes are used to verify the performance and effectiveness of the novel approach. Specifically, in comparison to the existing UAV deployment methods, such as random deployment and uniform deployment, the best overall performance is provided by the suggested MOEA/D-CICM. Moreover, it also has better performance in comparing to other MOEAs, such as MOCRY, MOPSO, MOFPA, and the conventional MOEA/D in solving the formulated UAVDMOP. In our upcoming work, we will take into account more realistic transmission conditions to further evaluate the effectiveness of the proposed method. To further assess the efficacy of the suggested technique, we will take into account more realistic transmission situations in our next study.

\section{}\label{}

\printcredits

\bibliographystyle{cas-model2-names}

\bibliography{ref-UAV-VLC}

\begin{thebibliography}{53}
\expandafter\ifx\csname natexlab\endcsname\relax\def\natexlab#1{#1}\fi
\providecommand{\url}[1]{\texttt{#1}}
\providecommand{\href}[2]{#2}
\providecommand{\path}[1]{#1}
\providecommand{\DOIprefix}{doi:}
\providecommand{\ArXivprefix}{arXiv:}
\providecommand{\URLprefix}{URL: }
\providecommand{\Pubmedprefix}{pmid:}
\providecommand{\doi}[1]{\href{http://dx.doi.org/#1}{\path{#1}}}
\providecommand{\Pubmed}[1]{\href{pmid:#1}{\path{#1}}}
\providecommand{\bibinfo}[2]{#2}
\ifx\xfnm\relax \def\xfnm[#1]{\unskip,\space#1}\fi
\bibitem[{Abbas et~al.(2018)Abbas, Song and Hong}]{abbas2018opportunistic}
\bibinfo{author}{Abbas, M.A.}, \bibinfo{author}{Song, H.}, \bibinfo{author}{Hong, J.P.}, \bibinfo{year}{2018}.
\newblock \bibinfo{title}{Opportunistic scheduling for average secrecy rate enhancement in fading downlink channel with potential eavesdroppers}.
\newblock \bibinfo{journal}{IEEE Trans. Inf. Forensics Secur} \bibinfo{volume}{14}, \bibinfo{pages}{969--980}.
\bibitem[{Akgun et~al.(2016)Akgun, Koyluoglu and Krunz}]{akgun2016exploiting}
\bibinfo{author}{Akgun, B.}, \bibinfo{author}{Koyluoglu, O.O.}, \bibinfo{author}{Krunz, M.}, \bibinfo{year}{2016}.
\newblock \bibinfo{title}{Exploiting full-duplex receivers for achieving secret communications in multiuser {MISO} networks}.
\newblock \bibinfo{journal}{IEEE Trans. Commun} \bibinfo{volume}{65}, \bibinfo{pages}{956--968}.
\bibitem[{Al-Khori et~al.(2018)Al-Khori, Nauryzbayev, Abdallah and Hamdi}]{al2018physical}
\bibinfo{author}{Al-Khori, J.}, \bibinfo{author}{Nauryzbayev, G.}, \bibinfo{author}{Abdallah, M.}, \bibinfo{author}{Hamdi, M.}, \bibinfo{year}{2018}.
\newblock \bibinfo{title}{Physical layer security for hybrid {RF/VLC} {DF} relaying systems}, in: \bibinfo{booktitle}{VTC-Fall}, \bibinfo{organization}{IEEE}. pp. \bibinfo{pages}{1--6}.
\bibitem[{Bai et~al.(2019)Bai, Yang, Guo, Feng and Xu}]{2019Camera}
\bibinfo{author}{Bai, L.}, \bibinfo{author}{Yang, Y.}, \bibinfo{author}{Guo, C.}, \bibinfo{author}{Feng, C.}, \bibinfo{author}{Xu, X.}, \bibinfo{year}{2019}.
\newblock \bibinfo{title}{Camera assisted received signal strength ratio algorithm for indoor visible light positioning}.
\newblock \bibinfo{journal}{IEEE Commun. Lett} \bibinfo{volume}{PP}, \bibinfo{pages}{1--1}.
\bibitem[{Chu et~al.(2015)Chu, Xing, Johnston and Le~Goff}]{chu2015secrecy}
\bibinfo{author}{Chu, Z.}, \bibinfo{author}{Xing, H.}, \bibinfo{author}{Johnston, M.}, \bibinfo{author}{Le~Goff, S.}, \bibinfo{year}{2015}.
\newblock \bibinfo{title}{Secrecy rate optimizations for a {MISO} secrecy channel with multiple multiantenna eavesdroppers}.
\newblock \bibinfo{journal}{IEEE Trans. Wirel. Commun} \bibinfo{volume}{15}, \bibinfo{pages}{283--297}.
\bibitem[{Coello and Lechuga(2002)}]{coello2002mopso}
\bibinfo{author}{Coello, C.C.}, \bibinfo{author}{Lechuga, M.S.}, \bibinfo{year}{2002}.
\newblock \bibinfo{title}{{MOPSO}: A proposal for multiple objective particle swarm optimization}, in: \bibinfo{booktitle}{Proceedings of the 2002 Congress on Evolutionary Computation. CEC'02 (Cat. No. 02TH8600)}, \bibinfo{organization}{IEEE}. pp. \bibinfo{pages}{1051--1056}.
\bibitem[{Deb et~al.(2002)Deb, Pratap, Agarwal and Meyarivan}]{deb2002fast}
\bibinfo{author}{Deb, K.}, \bibinfo{author}{Pratap, A.}, \bibinfo{author}{Agarwal, S.}, \bibinfo{author}{Meyarivan, T.}, \bibinfo{year}{2002}.
\newblock \bibinfo{title}{A fast and elitist multiobjective genetic algorithm: {NSGA-II}}.
\newblock \bibinfo{journal}{IEEE Trans. Evol. Comput} \bibinfo{volume}{6}, \bibinfo{pages}{182--197}.
\bibitem[{Dhanya and Kumar(2010)}]{dhanya2010nonlinear}
\bibinfo{author}{Dhanya, C.}, \bibinfo{author}{Kumar, D.N.}, \bibinfo{year}{2010}.
\newblock \bibinfo{title}{Nonlinear ensemble prediction of chaotic daily rainfall}.
\newblock \bibinfo{journal}{Advances in Water resources} \bibinfo{volume}{33}, \bibinfo{pages}{327--347}.
\bibitem[{Eltokhey et~al.(2021)Eltokhey, Khalighi and Ghassemlooy}]{eltokhey2021uav}
\bibinfo{author}{Eltokhey, M.W.}, \bibinfo{author}{Khalighi, M.A.}, \bibinfo{author}{Ghassemlooy, Z.}, \bibinfo{year}{2021}.
\newblock \bibinfo{title}{{UAV} location optimization in {MISO} {ZF} pre-coded {VLC} networks}.
\newblock \bibinfo{journal}{IEEE Wirel. Commun. Lett} \bibinfo{volume}{11}, \bibinfo{pages}{28--32}.
\bibitem[{Goel and Negi(2008)}]{goel2008guaranteeing}
\bibinfo{author}{Goel, S.}, \bibinfo{author}{Negi, R.}, \bibinfo{year}{2008}.
\newblock \bibinfo{title}{Guaranteeing secrecy using artificial noise}.
\newblock \bibinfo{journal}{IEEE Trans. Wirel. Commun} \bibinfo{volume}{7}, \bibinfo{pages}{2180--2189}.
\bibitem[{Hu et~al.(2017)Hu, Wen, Wu, Pan, Liao, Song, Tang and Wang}]{hu2017cooperative}
\bibinfo{author}{Hu, L.}, \bibinfo{author}{Wen, H.}, \bibinfo{author}{Wu, B.}, \bibinfo{author}{Pan, F.}, \bibinfo{author}{Liao, R.F.}, \bibinfo{author}{Song, H.}, \bibinfo{author}{Tang, J.}, \bibinfo{author}{Wang, X.}, \bibinfo{year}{2017}.
\newblock \bibinfo{title}{Cooperative jamming for physical layer security enhancement in {I}nternet of {T}hings}.
\newblock \bibinfo{journal}{IEEE Internet Things J} \bibinfo{volume}{5}, \bibinfo{pages}{219--228}.
\bibitem[{Huang et~al.(2018)Huang, Li, Liang, Xue and Wang}]{huang2018efficient}
\bibinfo{author}{Huang, Y.}, \bibinfo{author}{Li, W.}, \bibinfo{author}{Liang, Z.}, \bibinfo{author}{Xue, Y.}, \bibinfo{author}{Wang, X.}, \bibinfo{year}{2018}.
\newblock \bibinfo{title}{Efficient business process consolidation: combining topic features with structure matching}.
\newblock \bibinfo{journal}{Soft Comput} \bibinfo{volume}{22}, \bibinfo{pages}{645--657}.
\bibitem[{Khodadadi et~al.(2021)Khodadadi, Azizi, Talatahari and Sareh}]{khodadadi2021multi}
\bibinfo{author}{Khodadadi, N.}, \bibinfo{author}{Azizi, M.}, \bibinfo{author}{Talatahari, S.}, \bibinfo{author}{Sareh, P.}, \bibinfo{year}{2021}.
\newblock \bibinfo{title}{Multi-objective crystal structure algorithm ({MOC}ry{S}t{A}l): Introduction and performance evaluation}.
\newblock \bibinfo{journal}{IEEE Access} \bibinfo{volume}{9}, \bibinfo{pages}{117795--117812}.
\bibitem[{Komine and Nakagawa(2004)}]{komine2004fundamental}
\bibinfo{author}{Komine, T.}, \bibinfo{author}{Nakagawa, M.}, \bibinfo{year}{2004}.
\newblock \bibinfo{title}{Fundamental analysis for visible-light communication system using {LED} lights}.
\newblock \bibinfo{journal}{IEEE Trans. Consumer Electron} \bibinfo{volume}{50}, \bibinfo{pages}{100--107}.
\bibitem[{Kumar et~al.(2020)Kumar, Garg and Gupta}]{kumar2020pls}
\bibinfo{author}{Kumar, A.}, \bibinfo{author}{Garg, P.}, \bibinfo{author}{Gupta, A.}, \bibinfo{year}{2020}.
\newblock \bibinfo{title}{{PLS} analysis in an indoor heterogeneous {VLC/RF} network based on known and unknown {CSI}}.
\newblock \bibinfo{journal}{IEEE Syst. J} \bibinfo{volume}{15}, \bibinfo{pages}{68--76}.
\bibitem[{Lan et~al.(2020)Lan, Ren, Chen and Cai}]{lan2020achievable}
\bibinfo{author}{Lan, X.}, \bibinfo{author}{Ren, J.}, \bibinfo{author}{Chen, Q.}, \bibinfo{author}{Cai, L.}, \bibinfo{year}{2020}.
\newblock \bibinfo{title}{Achievable secrecy rate region for buffer-aided multiuser {MISO} systems}.
\newblock \bibinfo{journal}{IEEE Trans. Inf. Forensics Secur} \bibinfo{volume}{15}, \bibinfo{pages}{3311--3324}.
\bibitem[{Lee et~al.(2018)Lee, Hong, Choi and Levorato}]{lee2018adaptive}
\bibinfo{author}{Lee, K.}, \bibinfo{author}{Hong, J.P.}, \bibinfo{author}{Choi, H.H.}, \bibinfo{author}{Levorato, M.}, \bibinfo{year}{2018}.
\newblock \bibinfo{title}{Adaptive wireless-powered relaying schemes with cooperative jamming for two-hop secure communication}.
\newblock \bibinfo{journal}{IEEE Internet Things J} \bibinfo{volume}{5}, \bibinfo{pages}{2793--2803}.
\bibitem[{Li and Zhang(2008)}]{li2008multiobjective}
\bibinfo{author}{Li, H.}, \bibinfo{author}{Zhang, Q.}, \bibinfo{year}{2008}.
\newblock \bibinfo{title}{Multiobjective optimization problems with complicated {P}areto sets, {MOEA/D} and {NSGA-II}}.
\newblock \bibinfo{journal}{IEEE Trans. Evol. Comput} \bibinfo{volume}{13}, \bibinfo{pages}{284--302}.
\bibitem[{Liang et~al.(2020)Liang, Fang, Sun and Zhang}]{liang2020physical}
\bibinfo{author}{Liang, S.}, \bibinfo{author}{Fang, Z.}, \bibinfo{author}{Sun, G.}, \bibinfo{author}{Zhang, J.}, \bibinfo{year}{2020}.
\newblock \bibinfo{title}{A physical layer security approach based on optical beamforming for indoor visible light communication}.
\newblock \bibinfo{journal}{IEEE Commun. Lett} \bibinfo{volume}{24}, \bibinfo{pages}{2109--2113}.
\bibitem[{Liang et~al.(2009)Liang, Poor and Shamai}]{liang2009information}
\bibinfo{author}{Liang, Y.}, \bibinfo{author}{Poor, H.V.}, \bibinfo{author}{Shamai, S.}, \bibinfo{year}{2009}.
\newblock \bibinfo{title}{Information theoretic security}.
\newblock \bibinfo{publisher}{Now Publishers Inc}.
\bibitem[{Lin et~al.(2017a)Lin, Xu, He and Li}]{lin2017multi}
\bibinfo{author}{Lin, W.}, \bibinfo{author}{Xu, S.}, \bibinfo{author}{He, L.}, \bibinfo{author}{Li, J.}, \bibinfo{year}{2017}a.
\newblock \bibinfo{title}{Multi-resource scheduling and power simulation for cloud computing}.
\newblock \bibinfo{journal}{Inf. Sci} \bibinfo{volume}{397}, \bibinfo{pages}{168--186}.
\bibitem[{Lin et~al.(2017b)Lin, Xu, Li, Xu and Peng}]{lin2017design}
\bibinfo{author}{Lin, W.}, \bibinfo{author}{Xu, S.}, \bibinfo{author}{Li, J.}, \bibinfo{author}{Xu, L.}, \bibinfo{author}{Peng, Z.}, \bibinfo{year}{2017}b.
\newblock \bibinfo{title}{Design and theoretical analysis of virtual machine placement algorithm based on peak workload characteristics}.
\newblock \bibinfo{journal}{Soft Comput} \bibinfo{volume}{21}, \bibinfo{pages}{1301--1314}.
\bibitem[{Liu et~al.(2022)Liu, Wang, Sun and Li}]{liu2022multi}
\bibinfo{author}{Liu, L.}, \bibinfo{author}{Wang, A.}, \bibinfo{author}{Sun, G.}, \bibinfo{author}{Li, J.}, \bibinfo{year}{2022}.
\newblock \bibinfo{title}{Multi-objective optimization for improving throughput and energy efficiency in {UAV}-enabled {I}o{T}}.
\newblock \bibinfo{journal}{IEEE Internet Things J} .
\bibitem[{Liu et~al.(2020)Liu, Wang, Zhou, Ma, Hu and Ng}]{liu2020beamforming}
\bibinfo{author}{Liu, X.}, \bibinfo{author}{Wang, Y.}, \bibinfo{author}{Zhou, F.}, \bibinfo{author}{Ma, S.}, \bibinfo{author}{Hu, R.Q.}, \bibinfo{author}{Ng, D.W.K.}, \bibinfo{year}{2020}.
\newblock \bibinfo{title}{Beamforming design for secure {MISO} visible light communication networks with {SLIPT}}.
\newblock \bibinfo{journal}{IEEE Trans. Commun} \bibinfo{volume}{68}, \bibinfo{pages}{7795--7809}.
\bibitem[{Lorenz(1963)}]{lorenz1963deterministic}
\bibinfo{author}{Lorenz, E.N.}, \bibinfo{year}{1963}.
\newblock \bibinfo{title}{Deterministic nonperiodic flow}.
\newblock \bibinfo{journal}{Journal of atmospheric sciences} \bibinfo{volume}{20}, \bibinfo{pages}{130--141}.
\bibitem[{Mostafa and Lampe(2015)}]{mostafa2015physical}
\bibinfo{author}{Mostafa, A.}, \bibinfo{author}{Lampe, L.}, \bibinfo{year}{2015}.
\newblock \bibinfo{title}{Physical-layer security for {MISO} visible light communication channels}.
\newblock \bibinfo{journal}{IEEE J. Sel. Areas Commun} \bibinfo{volume}{33}, \bibinfo{pages}{1806--1818}.
\bibitem[{Mozaffari et~al.(2019)Mozaffari, Saad, Bennis, Nam and Debbah}]{mozaffari2019tutorial}
\bibinfo{author}{Mozaffari, M.}, \bibinfo{author}{Saad, W.}, \bibinfo{author}{Bennis, M.}, \bibinfo{author}{Nam, Y.H.}, \bibinfo{author}{Debbah, M.}, \bibinfo{year}{2019}.
\newblock \bibinfo{title}{A tutorial on {UAV}s for wireless networks: Applications, challenges, and open problems}.
\newblock \bibinfo{journal}{IEEE Commun. Surv. Tutorials} \bibinfo{volume}{21}, \bibinfo{pages}{2334--2360}.
\bibitem[{Mukherjee et~al.(2014)Mukherjee, Fakoorian, Huang and Swindlehurst}]{mukherjee2014principles}
\bibinfo{author}{Mukherjee, A.}, \bibinfo{author}{Fakoorian, S.A.A.}, \bibinfo{author}{Huang, J.}, \bibinfo{author}{Swindlehurst, A.L.}, \bibinfo{year}{2014}.
\newblock \bibinfo{title}{Principles of physical layer security in multiuser wireless networks: A survey}.
\newblock \bibinfo{journal}{IEEE Commun. Surv. Tutorials} \bibinfo{volume}{16}, \bibinfo{pages}{1550--1573}.
\bibitem[{Peer et~al.(2022)Peer, Lata, Srivastava, Bohara et~al.}]{peer20223}
\bibinfo{author}{Peer, M.}, \bibinfo{author}{Lata, K.}, \bibinfo{author}{Srivastava, A.}, \bibinfo{author}{Bohara, V.A.}, et~al., \bibinfo{year}{2022}.
\newblock \bibinfo{title}{3-{D} deployment of {VLC} enabled {UAV} networks with energy and user mobility awareness}.
\newblock \bibinfo{journal}{IEEE Transactions on Green Communications and Networking} , \bibinfo{pages}{1--1}.
\bibitem[{Peng et~al.(2021)Peng, Wang, Han and Jiang}]{peng2021physical}
\bibinfo{author}{Peng, H.}, \bibinfo{author}{Wang, Z.}, \bibinfo{author}{Han, S.}, \bibinfo{author}{Jiang, Y.}, \bibinfo{year}{2021}.
\newblock \bibinfo{title}{Physical layer security for {MISO} {NOMA} {VLC} system under eavesdropper collusion}.
\newblock \bibinfo{journal}{IEEE Trans. Veh. Technol} \bibinfo{volume}{70}, \bibinfo{pages}{6249--6254}.
\bibitem[{Pham et~al.(2020)Pham, Huynh-The, Alazab, Zhao and Hwang}]{pham2020sum}
\bibinfo{author}{Pham, Q.V.}, \bibinfo{author}{Huynh-The, T.}, \bibinfo{author}{Alazab, M.}, \bibinfo{author}{Zhao, J.}, \bibinfo{author}{Hwang, W.J.}, \bibinfo{year}{2020}.
\newblock \bibinfo{title}{Sum-rate maximization for {UAV}-assisted visible light communications using {NOMA}: Swarm intelligence meets machine learning}.
\newblock \bibinfo{journal}{IEEE Internet Things J} \bibinfo{volume}{7}, \bibinfo{pages}{10375--10387}.
\bibitem[{Saremi et~al.(2014)Saremi, Mirjalili and Lewis}]{saremi2014biogeography}
\bibinfo{author}{Saremi, S.}, \bibinfo{author}{Mirjalili, S.}, \bibinfo{author}{Lewis, A.}, \bibinfo{year}{2014}.
\newblock \bibinfo{title}{Biogeography-based optimisation with chaos}.
\newblock \bibinfo{journal}{Neural Comput. Appl} \bibinfo{volume}{25}, \bibinfo{pages}{1077--1097}.
\bibitem[{Sarveswararao et~al.(2022)Sarveswararao, Ravi and Huq}]{sarveswararao2022optimal}
\bibinfo{author}{Sarveswararao, V.}, \bibinfo{author}{Ravi, V.}, \bibinfo{author}{Huq, S.T.U.}, \bibinfo{year}{2022}.
\newblock \bibinfo{title}{Optimal prediction intervals for macroeconomic time series using chaos and evolutionary multi-objective optimization algorithms}.
\newblock \bibinfo{journal}{Swarm and Evolutionary Computation} \bibinfo{volume}{71}, \bibinfo{pages}{101070}.
\bibitem[{Shannon(1949)}]{shannon1949communication}
\bibinfo{author}{Shannon, C.E.}, \bibinfo{year}{1949}.
\newblock \bibinfo{title}{Communication theory of secrecy systems}.
\newblock \bibinfo{journal}{Bell Syst. Tech. J} \bibinfo{volume}{28}, \bibinfo{pages}{656--715}.
\bibitem[{Shiu et~al.(2011)Shiu, Chang, Wu, Huang and Chen}]{shiu2011physical}
\bibinfo{author}{Shiu, Y.S.}, \bibinfo{author}{Chang, S.Y.}, \bibinfo{author}{Wu, H.C.}, \bibinfo{author}{Huang, S.C.H.}, \bibinfo{author}{Chen, H.H.}, \bibinfo{year}{2011}.
\newblock \bibinfo{title}{Physical layer security in wireless networks: A tutorial}.
\newblock \bibinfo{journal}{IEEE Wirel. Commun} \bibinfo{volume}{18}, \bibinfo{pages}{66--74}.
\bibitem[{Sun et~al.(2021)Sun, Li, Liu, Liang and Kang}]{sun2021time}
\bibinfo{author}{Sun, G.}, \bibinfo{author}{Li, J.}, \bibinfo{author}{Liu, Y.}, \bibinfo{author}{Liang, S.}, \bibinfo{author}{Kang, H.}, \bibinfo{year}{2021}.
\newblock \bibinfo{title}{Time and energy minimization communications based on collaborative beamforming for {UAV} networks: A multi-objective optimization method}.
\newblock \bibinfo{journal}{IEEE J. Sel. Areas Commun} \bibinfo{volume}{39}, \bibinfo{pages}{3555--3572}.
\bibitem[{Syswerda et~al.(1989)}]{syswerda1989uniform}
\bibinfo{author}{Syswerda, G.}, et~al., \bibinfo{year}{1989}.
\newblock \bibinfo{title}{Uniform crossover in genetic algorithms.}, in: \bibinfo{booktitle}{ICGA}, pp. \bibinfo{pages}{2--9}.
\bibitem[{Tang et~al.(2018)Tang, Wang and Dong}]{tang2018adaptive}
\bibinfo{author}{Tang, L.}, \bibinfo{author}{Wang, X.}, \bibinfo{author}{Dong, Z.}, \bibinfo{year}{2018}.
\newblock \bibinfo{title}{Adaptive multiobjective differential evolution with reference axis vicinity mechanism}.
\newblock \bibinfo{journal}{IEEE Trans. Cybern} \bibinfo{volume}{49}, \bibinfo{pages}{3571--3585}.
\bibitem[{Trivedi et~al.(2016)Trivedi, Srinivasan, Sanyal and Ghosh}]{trivedi2016survey}
\bibinfo{author}{Trivedi, A.}, \bibinfo{author}{Srinivasan, D.}, \bibinfo{author}{Sanyal, K.}, \bibinfo{author}{Ghosh, A.}, \bibinfo{year}{2016}.
\newblock \bibinfo{title}{A survey of multiobjective evolutionary algorithms based on decomposition}.
\newblock \bibinfo{journal}{IEEE Trans. Evol. Comput} \bibinfo{volume}{21}, \bibinfo{pages}{440--462}.
\bibitem[{Vesterstrom and Thomsen(2004)}]{vesterstrom2004comparative}
\bibinfo{author}{Vesterstrom, J.}, \bibinfo{author}{Thomsen, R.}, \bibinfo{year}{2004}.
\newblock \bibinfo{title}{A comparative study of differential evolution, particle swarm optimization, and evolutionary algorithms on numerical benchmark problems}, in: \bibinfo{booktitle}{Proceedings of the 2004 congress on evolutionary computation (IEEE Cat. No. 04TH8753)}, \bibinfo{organization}{IEEE}. pp. \bibinfo{pages}{1980--1987}.
\bibitem[{Wang et~al.(2017)Wang, Teh and Li}]{wang2017artificial}
\bibinfo{author}{Wang, W.}, \bibinfo{author}{Teh, K.C.}, \bibinfo{author}{Li, K.H.}, \bibinfo{year}{2017}.
\newblock \bibinfo{title}{Artificial noise aided physical layer security in multi-antenna small-cell networks}.
\newblock \bibinfo{journal}{IEEE Trans. Inf. Forensics Secur} \bibinfo{volume}{12}, \bibinfo{pages}{1470--1482}.
\bibitem[{Wang et~al.(2020a)Wang, Chen, Yang, Luo and Saad}]{wang2020deep}
\bibinfo{author}{Wang, Y.}, \bibinfo{author}{Chen, M.}, \bibinfo{author}{Yang, Z.}, \bibinfo{author}{Luo, T.}, \bibinfo{author}{Saad, W.}, \bibinfo{year}{2020}a.
\newblock \bibinfo{title}{Deep learning for optimal deployment of {UAV}s with visible light communications}.
\newblock \bibinfo{journal}{IEEE Trans. Wirel. Commun} \bibinfo{volume}{19}, \bibinfo{pages}{7049--7063}.
\bibitem[{Wang et~al.(2020b)Wang, Dong, Hu and Wang}]{wang2020improved}
\bibinfo{author}{Wang, Y.}, \bibinfo{author}{Dong, Z.}, \bibinfo{author}{Hu, T.}, \bibinfo{author}{Wang, X.}, \bibinfo{year}{2020}b.
\newblock \bibinfo{title}{An improved {MOEA/D} algorithm for the carbon black production line static and dynamic multiobjective scheduling problem}, in: \bibinfo{booktitle}{2020 IEEE Congress on Evolutionary Computation (CEC)}, \bibinfo{organization}{IEEE}. pp. \bibinfo{pages}{1--8}.
\bibitem[{Wyner(1975)}]{wyner1975wire}
\bibinfo{author}{Wyner, A.D.}, \bibinfo{year}{1975}.
\newblock \bibinfo{title}{The wire-tap channel}.
\newblock \bibinfo{journal}{Bell system technical journal} \bibinfo{volume}{54}, \bibinfo{pages}{1355--1387}.
\bibitem[{Yang et~al.(2014)Yang, Karamanoglu and He}]{yang2014flower}
\bibinfo{author}{Yang, X.S.}, \bibinfo{author}{Karamanoglu, M.}, \bibinfo{author}{He, X.}, \bibinfo{year}{2014}.
\newblock \bibinfo{title}{Flower pollination algorithm: A novel approach for multiobjective optimization}.
\newblock \bibinfo{journal}{Engineering optimization} \bibinfo{volume}{46}, \bibinfo{pages}{1222--1237}.
\bibitem[{Yang et~al.(2019)Yang, Chen, Guo, Feng and Saad}]{yang2019power}
\bibinfo{author}{Yang, Y.}, \bibinfo{author}{Chen, M.}, \bibinfo{author}{Guo, C.}, \bibinfo{author}{Feng, C.}, \bibinfo{author}{Saad, W.}, \bibinfo{year}{2019}.
\newblock \bibinfo{title}{Power efficient visible light communication with unmanned aerial vehicles}.
\newblock \bibinfo{journal}{IEEE Commun. Lett.} \bibinfo{volume}{23}, \bibinfo{pages}{1272--1275}.
\bibitem[{Yanheng et~al.(2022)Yanheng, Lu, Sun and Liu}]{lu2022Optical}
\bibinfo{author}{Yanheng, L.}, \bibinfo{author}{Lu, J.}, \bibinfo{author}{Sun, G.}, \bibinfo{author}{Liu, L.}, \bibinfo{year}{2022}.
\newblock \bibinfo{title}{Optical power coverage optimization for {UAV}-enabled visible light communication}.
\newblock \bibinfo{journal}{International Conference on Communications (ICC)} .
\bibitem[{Zeng et~al.(2019)Zeng, Xu and Zhang}]{2019Energy}
\bibinfo{author}{Zeng, Y.}, \bibinfo{author}{Xu, J.}, \bibinfo{author}{Zhang, R.}, \bibinfo{year}{2019}.
\newblock \bibinfo{title}{Energy minimization for wireless communication with {R}otary-wing {UAV}}.
\newblock \bibinfo{journal}{IEEE Trans. Wirel. Commun} \bibinfo{volume}{18}, \bibinfo{pages}{2329--2345}.
\bibitem[{Zhang and Li(2007)}]{zhang2007moea}
\bibinfo{author}{Zhang, Q.}, \bibinfo{author}{Li, H.}, \bibinfo{year}{2007}.
\newblock \bibinfo{title}{{MOEA/D}: A multiobjective evolutionary algorithm based on decomposition}.
\newblock \bibinfo{journal}{IEEE Trans. Evol. Comput} \bibinfo{volume}{11}, \bibinfo{pages}{712--731}.
\bibitem[{Zhang et~al.(2009)Zhang, Hu, Wang and Zhu}]{zhang2009application}
\bibinfo{author}{Zhang, S.}, \bibinfo{author}{Hu, Q.}, \bibinfo{author}{Wang, X.}, \bibinfo{author}{Zhu, Z.}, \bibinfo{year}{2009}.
\newblock \bibinfo{title}{Application of chaos genetic algorithm to transformer optimal design}, in: \bibinfo{booktitle}{2009 International Workshop on Chaos-Fractals Theories and Applications}, \bibinfo{organization}{IEEE}. pp. \bibinfo{pages}{108--111}.
\bibitem[{Zhao et~al.(2012)Zhao, Suganthan and Zhang}]{zhao2012decomposition}
\bibinfo{author}{Zhao, S.Z.}, \bibinfo{author}{Suganthan, P.N.}, \bibinfo{author}{Zhang, Q.}, \bibinfo{year}{2012}.
\newblock \bibinfo{title}{Decomposition-based multiobjective evolutionary algorithm with an ensemble of neighborhood sizes}.
\newblock \bibinfo{journal}{IEEE Trans. Evol. Comput} \bibinfo{volume}{16}, \bibinfo{pages}{442--446}.
\bibitem[{Zheng et~al.(2018)Zheng, Tan, Meng and Zhang}]{zheng2018improved}
\bibinfo{author}{Zheng, W.}, \bibinfo{author}{Tan, Y.}, \bibinfo{author}{Meng, L.}, \bibinfo{author}{Zhang, H.}, \bibinfo{year}{2018}.
\newblock \bibinfo{title}{An improved {MOEA}/{D} design for many-objective optimization problems}.
\newblock \bibinfo{journal}{Appl. Intell} \bibinfo{volume}{48}, \bibinfo{pages}{3839--3861}.
\bibitem[{Zhou et~al.(2019)Zhou, Wang, Xu, Yan and Zhu}]{zhou2019novel}
\bibinfo{author}{Zhou, J.}, \bibinfo{author}{Wang, F.}, \bibinfo{author}{Xu, J.}, \bibinfo{author}{Yan, Y.}, \bibinfo{author}{Zhu, H.}, \bibinfo{year}{2019}.
\newblock \bibinfo{title}{A novel character segmentation method for serial number on banknotes with complex background}.
\newblock \bibinfo{journal}{J. Ambient Intell. Humaniz. Comput} \bibinfo{volume}{10}, \bibinfo{pages}{2955--2969}.

\end{thebibliography}

\end{document}